\documentclass[english,dvips]{emulateapj}

\usepackage{color}
\usepackage{amssymb}
\usepackage{amssymb}
\usepackage{natbib}
\usepackage{amsmath}
\usepackage{graphicx}

\makeatletter

\providecommand{\tabularnewline}{\\}
\slugcomment{}
\shorttitle{}
\shortauthors{}
\providecommand{\boldsymbol}[1]{\mbox{\boldmath $#1$}}

\newcommand{\Kelvin}{\mathrm{K}}
\newcommand{\Msun}{\mathrm{M_{\sun}}}

\newcommand{\Lsun}{L_{\sun}}
\newcommand{\MsunPerYear}{\mathrm{M_{\sun}\,yr^{-1}}}

\makeatother

\begin{document}

\title{Three-Dimensional Simulations of Dynamics of Accretion Flows Irradiated
by a Quasar }

\author{Ryuichi Kurosawa and Daniel Proga}

\affil{Department of Physics and Astronomy, University of Nevada Las Vegas,
Box~454002, 4505~Maryland Pkwy, Las Vegas, NV 891541-4002}

\email{\{rk,dproga\}@physics.unlv.edu}

\begin{abstract}
We study the axisymmetric and non-axisymmetric, time-dependent
hydrodynamics of gas that is under the influence of the gravity of a
super massive black hole (SMBH) and the radiation force produced by a
radiatively efficient flow accreting onto the SMBH. We have considered
two cases: (1)~the formation of an outflow from the accretion of the
ambient gas without rotation and (2)~that with \textcolor{black}{weak} rotation. The main
goals of this study are: (1)~to examine if there is a significant
difference between the models with identical initial and boundary
conditions but in different dimensionality (2-D and 3-D), and (2)~to
understand the gas dynamics in AGN.  Our 3-D simulations of a
non-rotating gas show small yet noticeable non-axisymmetric
small-scale features inside the outflow.  The outflow as a whole and
the inflow do not seem to suffer from any large-scale instability. In
the rotating case, the non-axisymmetric features are very prominent,
especially in the outflow which consists of many cold dense clouds
entrained in a smoother hot flow. The 3-D outflow is
non-axisymmetric due to the shear and thermal instabilities. In both
2-D and 3-D simulations, gas rotation increases the outflow thermal
energy flux, but reduces the outflow mass and kinetic energy
fluxes. Rotation also leads to time variability and fragmentation of
the outflow in the radial and latitudinal directions. The collimation
of the outflow is reduced in the models with gas rotation. The time
variability in the mass and energy fluxes is reduced in the 3-D case
because of the outflow fragmentation in the azimuthal direction.  The
virial mass estimated from the kinematics of the dense cold clouds
found in our 3-D simulations of rotating gas underestimates the actual
mass used in the simulations by about 40~\%.  The opening angles
($\sim30^{\circ}$) of the bi-conic outflows found in the models with
rotating gas are very similar to that of the nearby Seyfert galaxy
NGC~4151 ($\sim33^{\circ}$).  The radial velocities of the dense cold
clouds from the simulations are compared with the observed gas
kinematics of the narrow line region of NGC~4151.
\end{abstract}


\section{Introduction}

\label{sec:Introduction}

Active Galactic Nuclei (AGNs) are powered by accretion of matter onto
a super massive ($10^{6}$--$10^{10}\,\Msun$) black hole (SMBH), and produce a large
amount of energy (e.g., \citealt{Lynden-Bell:1969}) as electromagnetic
radiation ($10^{10}$--$10^{14}\Lsun$), over a wide range of wavelengths
(from radio to hard X-rays, and even to TeV photons).  The strong radiation from AGNs
influences the physical properties (e.g., the ionization structure,
gas dynamics and density distribution) of their vicinity, their host
galaxies, and even of the inter-galactic material of galaxy clusters
to which they belong (e.g., \citealt*{Quilis:2001};
\citealt{DallaVecchia:2004}; \citealt{McNamara:2005};
\citealt{Zanni:2005}; \citealt{Fabian:2006}; \citealt{Vernaleo:2006}).
The importance of the radiation-driven outflows from AGNs as a
feedback process is recognized in many of the galaxy
formation/evolutionary models (e.g., \citealt{ciotti:1997},
\citeyear{ciotti:2001}, \citeyear{ciotti:2007}; \citealt{king:2003};
\citealt{Hopkins:2005}; \citealt*{Murray:2005};
\citealt{Sazonov:2005}; \citealt*{Springel:2005};
\citealt{Brighenti:2006}; \citealt{Fontanot:2006};
\citealt*{Wang:2006}, \citealt*{Tremonti:2007}; Ciotti et al.~2008, in
preparation).

The formation of AGN outflows, of course, can be caused by some 
mechanisms other than radiation pressure, e.g., magnetocentrifugal force (e.g.,
\citealt{Blandford:1982}; \citealt*{Emmering:1992};
\citealt{Konigl:1994}; \citealt{Bottorff:1997}),
\textcolor{black}{
Poynting flux/magnetic towers (e.g.,
\citealt{Lovelace:1987}; \citealt{Lynden-Bell:1996,Lynden-Bell:2003}; \citealt{Li:2001};
\citealt{Kato:2004}; \citealt{Nakamura:2006}; \citealt{Kato:2007}), 
}
 and thermal pressure
(e.g., \citealt{Weymann:1982}; \citealt*{Begelman:1991};
\citealt{Everett:2007}).  However, the highly blueshifted line
absorption features often seen in the observed UV and optical spectra
of AGNs can be best described by the radiation-driven wind models
(e.g.,~\citealt{Murray:1995b}; \citealt{Proga:2000};
\citealt{Proga:2004}), provided that the ionization state of the gas
is appropriate.  In reality, these forces may interplay and contribute
to the dynamics of the outflows in AGNs in somewhat different degrees
(e.g., \citealt{Konigl:2006}; \citealt{Proga:2007a}, and references
therein). 

The AGN environment on relatively large scales ($10^{2}-10^{3}$~pc)
is a mixture of gas and dust (e.g.~\citealt{Antonucci:1984};
\citealt{Miller:1990}; \citealt{Awaki:1991}; \citealt{Blanco:1990};
\citealt{Krolik:1999}).  The radiation pressure on dust can drive the
dust outflows, and their dynamics is likely to be coupled with the gas
dynamics (e.g., \citealt{Phinney:1989}; \citealt{Pier:1992};
\citealt{Emmering:1992}; \citealt{Laor:1993}; \citealt{Konigl:1994};
\citealt{Murray:2005}). On much smaller scales ($<\sim10$~pc), 
the dust is less likely to survive because the temperature
of the environment is too high ($>10^{4}\,\Kelvin$); hence, the studies
of the radiation-driven outflow dynamics using only gas component
(e.g.,~\citealt*{Arav:1994}; \citealt{Proga:2000}) would be justified in
those cases. 

In the first paper of this series (\citealt{Proga:2007b}, hereafter
Paper~I), the initial phase of our gas dynamics studies of AGNs on
sub-parsec and parsec scales was set.  Since the problem is complex,
as it involves many aspects of physics such as multi-dimensional fluid
dynamics, radiative processes, and magnetic processes, our approach
was to set up simulations as simple as possible. The study focused on
exploring the effects of X-ray heating (which is important in the
so-called preheated accretion; e.g., \citealt{Ostriker:1976};
\citealt{Park:2001, Park:2007}) and radiation pressure on gas that is 
gravitationally captured by a black hole (BH). We adopted the
numerical methods developed by \citet{Proga:2000} for studying
radiation-driven disk winds in AGNs. Our simulations covered a
relatively unexplored range of the distance from the central BH, i.e.,
the outer boundary of our simulations coincides with the
inner boundary of many galaxy models (e.g., \citealt{Springel:2005};
\citealt{ciotti:2007}), and our inner boundary starts just outside 
of the outer boundary of many BH accretion models (e.g.,
\citealt{Hawley:2002}; \citealt{Ohsuga:2007}). The effect of gas
rotation was not included in Paper~I.

In the second paper in this series (\citealt{Proga:2008}, hereafter
Paper~II), the effect of gas rotation, position-dependent radiation
temperature, density at large radii, and uniform X-ray background
radiation were explored. As in the non-rotating case considered in
Paper~I, the rotating flow settles into a configuration with two
components: (1)~an equatorial inflow and (2)~a bipolar inflow/outflow
with the outflow leaving the system along the polar axis. However, with
rotation the flow does not always reach a steady state.  In addition,
rotation reduces the outflow collimation and the outward fluxes of mass
and kinetic energy. Moreover rotation increases the outward flux of
the thermal energy, and it can lead to fragmentation and
time-variability of the outflow. It is also shown that the 
position-dependent radiation temperature can significantly change the
flow solution, i.e.,  the inflow in the equatorial region can
be replaced by a thermally driven outflow. As it \textcolor{black}{has} been
discussed and shown in the past (e.g., \citealt{ciotti:2007}; Ciotti et al.~2008, in
preparation), the self-consistently determined preheating/cooling from the
quasar radiation can significantly reduce the mass accretion rate of the
central BH. Our results clearly demonstrated that quasar radiation can
drive non-spherical, multi-temperature and very dynamic flows. This 
effect becomes dominant for the systems with luminosity in excess of
0.01 times the Eddington luminosity.

The work presented here is a direct extension of the previous
axi-symmetric models of Paper~I and Paper~II to a full 3-D model, and
is an extended version of the 3-D models presented in
\citet{Kurosawa:2008} to which we have added the radiation force due
to spectral lines and the radiative cooling and heating effect.  Here,
we consider two cases from Paper~I and Paper~II:
(1)~the formation of \textcolor{black}{relatively large scale ($\sim
  10$pc)} outflows from the accretion of the ambient
gas with no rotation and (2)~that with rotation, in 3-D.
We note that our work is complimentary to the work by
\citet{Dorodnitsyn:2008a, Dorodnitsyn:2008b} who studied the
hydrodynamics of axisymmetric torus winds in AGNs.

The main goals of this study are (1)~to examine if there is a
significant difference between two models with physically identical
conditions but in different dimensionality (2-D and 3-D), (2)~to study
if the radiation driven outflows that were found to be stable in the
previous studies in 2-D (Paper~I; Paper~II) remain stable in 3-D
simulations, and (3)~to understand gas dynamics in AGNs, in particular
the dynamics of the narrow line regions (NLR) by comparing our
simulation results with observations.

In the following section, we describe our method and model assumptions.
We give the results of our hydrodynamical simulations in \S~\ref{sec:Results}.
Discussions on virial mass estimates and comparisons with the observations
of Seyfert galaxies will be given in \S~\ref{sec:Discussions}.
The summary and conclusions are in \S~\ref{sec:Conclusions}.

\section{Method}
\label{sec:Method}

\subsection{Overview}
\label{sub:Overview}

\begin{figure}
\begin{center}

\includegraphics[clip,width=0.48\textwidth]{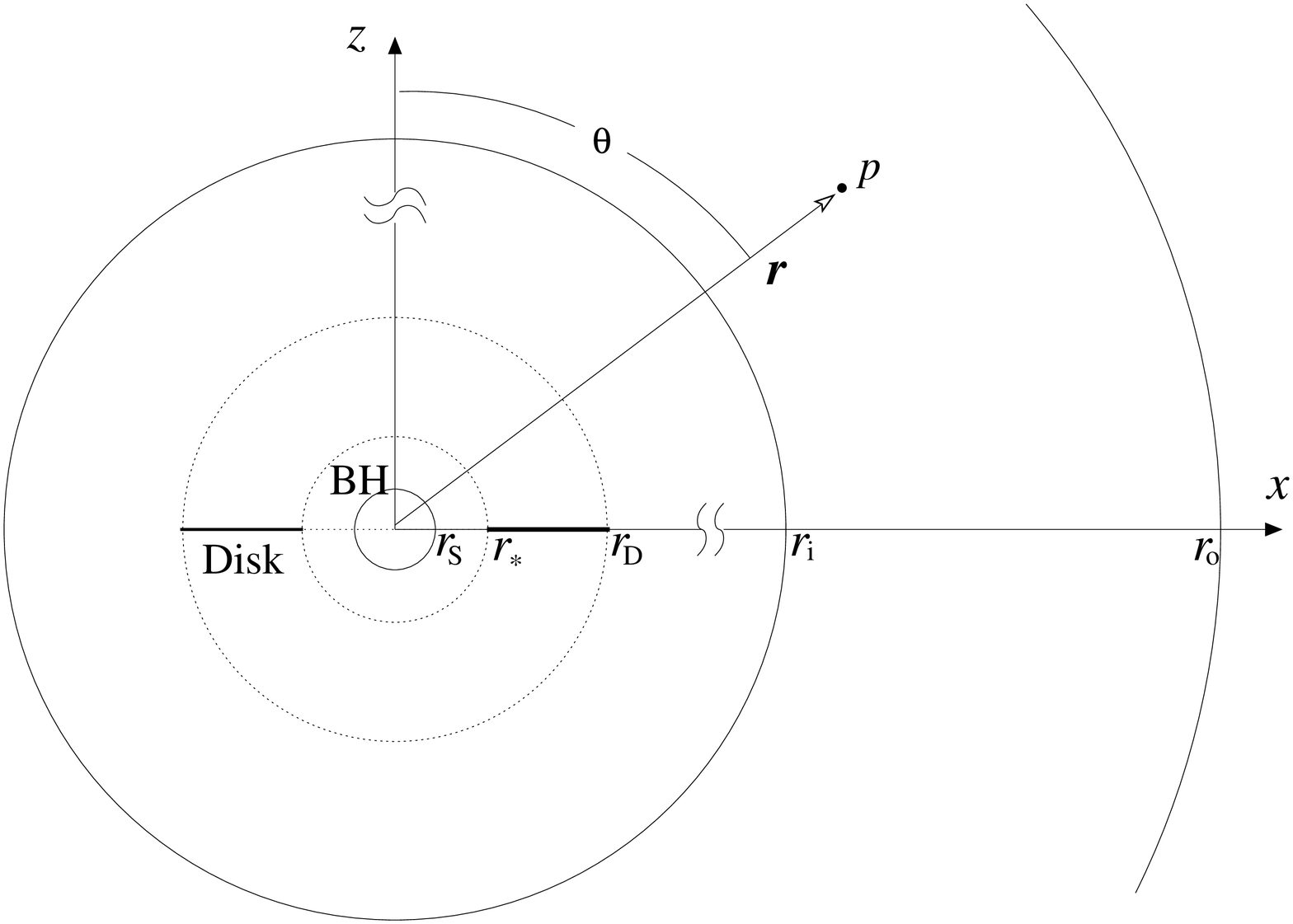}

\end{center}

\caption{Basic model configuration. In 3-D models, a super massive
  black hole
(BH) with its Schwarzschild radius $r_{S}$ is located at the center
of the cartesian coordinate system ($x$, $y$, $z$) where the $y$-axis
is perpendicular to and into the paper. The accretion disk spans from
its inner radius $r_{*}$ to its outer radius $r_{D}$. The 3-D hydrodynamic
simulations are performed in the spherical coordinate system ($r$,
$\theta$, $\phi$), and with $r$ between the inner
boundary $r_{\mathrm{i}}$ and the outer boundary $r_{\mathrm{o}}$.
For 2-D models, computations are performed on the $\phi=0$ plane
assuming an axisymmetry around the $z$-axis. While the radiation
pressure from the central BH on a point $p$ with its position vector
$\boldsymbol{r}$ is in radial direction and is function of $r$,
that from the accretion disk is assumed to be a function of $r$ and
$\theta$. A point-source approximation for the disk radiation force at
$p$ is valid when $r_{\mathrm{i}} \gg r_{\mathrm{D}}$. Note that
the figure is not to scale. }

\label{fig:model-config}
\end{figure}

We mainly follow the methods used \textcolor{black}{in} the axisymmetric models by \citet{Proga:2000}
and \citet{Proga:2004}, and extend the problems to a full 3-D. Our
basic model configuration is shown in Figure~\ref{fig:model-config}. 
The model geometry and the assumptions of the SMBH and the disk are
very similar to those in Paper~I, Paper~II
and \citet{Kurosawa:2008}. For the simulations in 3-D, a SMBH with
its mass $M_{\mathrm{BH}}$ and its Schwarzschild radius $r_{\mathrm{S}}=2GM_{\mathrm{BH}}/c^{2}$
is placed at the center of the spherical coordinate system ($r$,
$\theta$, $\phi$). The X-ray emitting corona regions is defined as a sphere
with its radius $r_{*}$, as shown in the figure. The geometrically
thin and optically thick accretion disk (e.g., \citealt{shakura:1973})
is placed on the equatorial plane ($\theta=\pi/2$ plane).
The 3-D hydrodynamic simulations will be performed in the spherical
coordinate system  with $r$ between the
inner boundary $r_{\mathrm{i}}$ and the outer boundary $r_{\mathrm{o}}$.
For 2-D models, the $z$-axis in the figure becomes the symmetry axis,
and the computations are performed on $\phi=0$ plane. The radiation
forces, from the corona region (the sphere with its radius $r_{*}$)
and the accretion disk, acting on the gas located at a location ($p$)
are assumed to be only in radial direction. The magnitude 
of the radiation force due to the corona is assumed to be a function
of radius $r$ only, but that due to the accretion disk is assumed to be
a function of $r$ and the polar angle $\theta$ which is the angle
between the $z$-axis and the position vector $\boldsymbol{r}$ as
shown in the figure. 
The point-source like approximation for the disk radiation pressure
at $p$ is used here since the accretion disk radius ($r_{\mathrm{D}}$
in Fig.~\ref{fig:model-config}) is assumed to be much smaller than the inner radius,
i.e., $r_{\mathrm{D}}\ll r_{\mathrm{i}}$.  
In the following, we will describe our radiation hydrodynamics,
our implementation of the radiation sources (the corona and disk),
and radiative cooling/heating. Finally, we will also describe the
model parameters and assumptions. 



%


\subsection{Hydrodynamics}

\label{sub:Hydrodynamics}

We employ 3-D hydrodynamical simulations of the outflow from and accretion
onto a central part of AGN, using the ZEUS-MP code \citep[c.f.,][]{Hayes:2006}
which is a massively parallel MPI-implemented version of the ZEUS-3D
code (c.f., \citealt{Hardee:1992}; \citealt{Clarke:1996}). The ZEUS-MP
is a Eulerian hydrodynamics code which uses the method of finite differencing
on a staggered mesh with a second-order-accurate, monotonic advection
scheme \citep{Hayes:2006}. To compute the structure and evolution
of a flow irradiated by a strong continuum radiation of AGN, we solve
the following set of HD equations: 
\begin{eqnarray}
  \frac{D\rho}{Dt}+\rho\,\boldsymbol{\nabla}\cdot\boldsymbol{v} & = & 0,
  \label{eq:hydro01}
\end{eqnarray}

\begin{equation}
  \rho\frac{D\boldsymbol{v}}{Dt}=-\boldsymbol{\nabla}P+\rho\,\boldsymbol{g}+\rho\,\boldsymbol{g}_{\mathrm{rad}},
  \label{eq:hydro02}
\end{equation}

\begin{equation}
  \rho\frac{D}{Dt}\left(\frac{e}{\rho}\right)=-P\,\boldsymbol{\nabla}\cdot\boldsymbol{v}+\rho\,\mathcal{C},
  \label{eq:hydro03}
\end{equation}
where $\rho$, $e$, $P$ and $\boldsymbol{v}$ are the mass density,
energy density, pressure, and the velocity of gas respectively. Also,
$\boldsymbol{g}$ is the gravitational force per unit mass. The Lagrangian/co-moving
derivative is defined as $D/Dt\equiv\partial/\partial t+\boldsymbol{v}\cdot\boldsymbol{\nabla}$.
We have introduced two new components to the ZEUS-MP in order to treat
the gas dynamics more appropriate for the gas flow in and around AGN.
The first is the acceleration due to radiative force per unit mass
($\boldsymbol{g}_{\mathrm{rad}}$) in equation~(\ref{eq:hydro02}),
and the second is the the effect of radiative cooling and heating
simply as the net cooling rate ($\mathcal{C}$) in equation~(\ref{eq:hydro03}).
In our previous 3-D models (\citealt{Kurosawa:2008}), we considered
a \textcolor{black}{simpler} case with $\mathcal{C}=0$, but here we generalize the
problem and consider cases with $\mathcal{C}\neq0$. We assume the
equation of state to be in the form of $P=\left(\gamma-1\right)e$
where $\gamma$ is the adiabatic index, and $\gamma=5/3$ for all
the models presented in this paper. 
Our numerical methods used in this paper are identical to, 
in most aspects,  \textcolor{black}{those described} in Paper~I and Paper II. In the
following, we describe only the key elements of the
calculations. Readers are referred to Paper~I and Paper~II (see also
\citealt{Proga:2000}) for details. 

Because of the accretion disk geometry (flat) which irradiates the
surrounding gas, the flows in our models will not be spherically
symmetric.  The disk radiation flux, ${\cal F}_{\rm disk}$ peaks in
the direction of the disk rotational axis, and it gradually decreases
as the polar angle $\theta$ increases, i.e., ${\cal F}_{\rm
  disk}\propto|\cos{\theta}|$.  The flow is also irradiated by a
corona which is assumed to be spherical.  The gas is assumed to be
optically thin to its own cooling radiation. The following radiative
processes are considered: Compton heating/cooling, X-ray
photoionization heating, and recombination, bremsstrahlung and line
cooling.  We take into account some effects of photoionization on
radiation pressure due to lines (line force).  The line force is
computed from a value of the photoionization parameter (defined as
$\xi=4\pi\mathcal{F_{\mathrm{X}}}/n$ where $\mathcal{F_{\mathrm{X}}}$
and $n$ are the local X-ray flux and the number density of the gas) in
combination with the analytical formulae from \citet{Stevens:1990}.
The attenuation of the X-ray radiation by computing the X-ray optical
depth in the radial direction is included.  On the other hand, we do
not include the attenuation of the UV radiation, to be consistent with
our gas heating rates in which we include the X-ray photoionization
but not UV photoionization. The method described above is found to be
computationally efficient (cf. Paper I and Paper II), and provides
good estimates for the number and opacity distribution of spectral
lines for a given $\xi$ without detail information about the
ionization state (see \citealt{Stevens:1990}).

Further, we assume that the total accretion luminosity $L$
\textcolor{black}{consists} 
of two components: (1)~$L_{\rm disk}=f_{\rm disk} L$ due to the
accretion disk and (2)~$L_\ast=f_{\rm \ast} L$ due to the corona.  We
assume that the disk emits only UV photons, whereas the corona emits
only X-rays, i.e., the system UV luminosity, $L_{\rm UV}=f_{\rm UV}
L=L_{\rm disk}$ and the system X-ray luminosity, $L_{\rm X}=f_{\rm X}
L=L_\ast$ (in other words $f_{\rm UV}=f_{\rm disk}$ and $f_{\rm
  X}=f_\ast$).

With these simplifications, only the corona radiation is responsible
for ionizing the flow to a very high ionization state. While the
corona contributes to the radiation force due to electron scattering
in our calculations, it does not contribute to line driving.  Metal
lines in the soft X-ray band may have an appreciable contribution to
the total radiation force in some cases.  The disk radiation
contributes to the radiation force due to both electron and line
scattering.

\subsection{Gas Rotation}
\label{sub:Gas-Rotation}

\textcolor{black}{For the simulations with gas rotation, we consider
  the accretion of gas with low specific angular momentum ($l$).  The
  low $l$ here means that the centrifugal force at large radii is small 
  compared to gravity and gas pressure. Thus, at large radii and
  without radiation pressure, the flow is almost radial.  However, at
  small radii, the flow starts to 
  converge toward the equator, and it can eventually form a
  rotation--pressure supported torus like ones studied by e.g.,
  \citet{Proga:2003c} (in 2-D) and \citet{Janiuk:2008} (in 3-D).  In
  general, gas at large radii would have a range of $l$, and
  some fraction of gas would converge toward the equator even at large
  radii. On the other hand, some fraction of gas would have very small
  $l$, and would directly fall onto the BH without going through a
  torus. }

\textcolor{black}{Following \citet{Proga:2003c} and Paper~II, we assume
that the initial distribution of specific angular momentum $l$, as a
function of the polar angle $\theta$, is}
\begin{equation}
  l\left(\theta\right)=l_{0}\, f\left(\theta\right),
  \label{eq:specific-angular-momentum}
\end{equation}
where $l_{0}$ is the specific angular momentum on the equator, and
$f\left(\theta\right)$ is a function monotonically decreases from
$1$ to $0$ from the equator to the poles (at $\theta=0^{\circ}$
and $180^{\circ}$). Using the {}``circularization radius'' $r'_{c}$
(in the units of $r_{*}$) 
\textcolor{black}{on the equator} 
for the Newtonian potential (i.e., $GM/r^{2}=v_{\phi}^{2}/r$
at $r=r'_{c}r_{*}$), the specific angular momentum on the equator
can be written as: 
\begin{equation}
  l_{0}=c\, r_{*}\sqrt{{\rm r'_{{\rm c}}/6}}
  \label{eq:L0}
\end{equation}
where $r_{*}=3r_{\mathrm{s}}=6\, GM/c^{2}$ is used. The angular dependency
in equation~(\ref{eq:specific-angular-momentum}) is chosen as:
\begin{equation}
  f(\theta)=1-|\cos\theta|.
  \label{eq:theta_func}
\end{equation}
The initial rotational velocity ($v_{0\phi}$) for the simulations
are assigned as:
\begin{equation}
  v_{0\phi}(r,\theta)=\left\{ 
  \begin{array}{ll}
    0 & \mathrm{for}\,\,\,\, r<10^{5}r_{*}\,,\\
    l/\sin\theta\, r & \mathrm{for}\,\,\,\, r\ge10^{5}r_{*}\,.
  \end{array}\right.
  \label{eq:v0phi}
\end{equation}

\textcolor{black}{
In this paper, we set $r'_{c}=300$ which is smaller than the inner
boundary radius ($r_\mathrm{i}=500\,r_{*}$).  This yields very
weakly rotating gas which is far from a rotational equilibrium inside
our computational domain. For example, the ratio of the centrifugal
acceleration to the gravitational acceleration on the equator at the
outer boundary ($r_\mathrm{o}=2.5\times10^{5}\,r_{*}$) is only $1.2\times
10^{-3}$.  We choose the relatively small value of $r'_{c}$ to avoid a
formation of a rotationally supported torus or disk in our
computational domain and to avoid the complexities associated with it, e.g.,
the instability (in non-axisymmetric modes) of a torus found by
\citet{Papaloizou:1984}.  The low value of
the gas specific angular momentum considered here allows us to study 
relatively simple flows, and to set an initial stage for modeling more
complex flows associated with larger values of specific angular
momentum, which shall be considered in a future study.
}

\textcolor{black}{
We assume that the circularized
gas, which would be formed at $r<r_\mathrm{i}$ (interior to the inner
radius of our
computational domain), will eventually accrete onto the SMBH on a
viscous timescale. We do not model the actual process(es) of the
angular momentum transport.  A most likely mechanism of the angular
momentum transport is magneto-rotational instability
(\citealt{Balbus:1991}).
}

\textcolor{black}{
The formation of a torus wind, which
might be associated with the X-ray ``warm absorbers'' (e.g.,
\citealt{Lira:1999}; \citealt{Moran:1999}; \citealt{Iwasawa:2000};
\citealt{Crenshaw:2004b}; \citealt{Blustin:2005}) in  
Seyfert galaxies, are considered elsewhere (e.g., \citealt{Dorodnitsyn:2008a,
  Dorodnitsyn:2008b}). Here we are interested in a lager scale ($\sim
10$~pc) weakly rotating wind  which might be relevant to the NLR
of AGNs.  Readers are refer to Paper~II for the axi-symmetric models
with a different choice of the specific angular distribution
function. 
}

\subsection{Model Setup}

\label{sub:Model-Setup}

In all models presented here, the following ranges of the coordinates
are adopted: $r_{\mathrm{i}}\leq r\leq r_{\mathrm{o}}$,
$0\leq\theta\leq\pi$ and $0\leq\phi<2\pi$ (for 3-D models) where
$r_{\mathrm{i}}=500\, r_{*}$ and $r_{\mathrm{o}}=2.5\times10^{5}\,
r_{*}$. The polar and azimuthal angle ranges are divided into 128 and
64 zones, and are equally spaced. In the $r$ direction, the gird is
divided into 128 zones in which the zone size ratio is fixed at
$\Delta r_{k+1}/\Delta r_{k}=1.04$.

For the initial conditions, the density and the temperature of gas
are set uniformly, i.e., $\rho=\rho_{o}$ and $T=T_{o}$ everywhere
in the computational domain where $\rho_{o}=1.0\times10^{-21}\,\mathrm{g\, cm^{-3}}$
and $T_{o}=2\times10^{7}\,\Kelvin$ \textcolor{black}{throughout} this paper (cf.~Paper~II).
For the models \textcolor{black}{without} gas rotation, the initial velocity is set to zero everywhere. 
For the models with gas rotation, the initial velocity of the gas is
assigned as \textcolor{black}{described in \S~\ref{sub:Gas-Rotation}} (see
also Paper~II).

At the inner and outer boundaries, we apply the outflow (free-to-outflow)
boundary conditions, in which the field values are extrapolated beyond
the boundaries using the values of \emph{the ghost zones} residing
outside of normal computational zones (see \citealt{Stone:1992} for
more details). At the outer boundary, all HD quantities (except the
radial component of the velocity, $v_{r}$) are assigned to the initial
conditions (e.g., $T=T_{o}$ and $\rho=\rho_{o}$) during the the
evolution of each model; however, this outer boundary condition is
applied only when the gas is inflowing at the outer boundary, i.e.,
when $v_{r}<0$. The radial component of the velocity is allowed to
float (unconstrained) when $v_{r}>0$ at the outer boundary. For the
models without gas rotation, $v_{\phi}=0$ is used for the outer boundary
condition while equation~(\ref{eq:v0phi}) is used for those with
gas rotation. Paper~II also applied these conditions to
represent a steady flow condition at the outer boundary. They found
that this technique leads to a solution that relaxes to a steady state
in both spherical and non-spherical accretion with an outflow (see
also \citealt{Proga:2003b}). This imitates the condition in which
a continuous supply of gas is available at the outer boundary.

\section{Results}

\label{sec:Results}

We consider models with and without gas rotation in both 2-D
and 3-D. The 2-D models are equivalent to Run~C (without rotation)
and Cr (with rotation) presented in Paper~I and Paper~II,
but here we used the newly modified 3-D version of the code
(ZEUS-MP). The 3-D models are equivalent to our 2-D models, but in 
those models, the assumption of the axisymmetry are dropped. We
examine the differences and similarities of the 2-D and 3-D models,
and investigate the importance of the non-axisymmetric natures of
the flows in 3-D. The main parameters and results of the four models
are summarized in Table~\ref{tab:Model-Summary}. In the following,
we describe the models results in detail.

\begin{figure*}
\begin{center}

\begin{tabular}{cc}
  \includegraphics[clip,height=0.35\textheight]{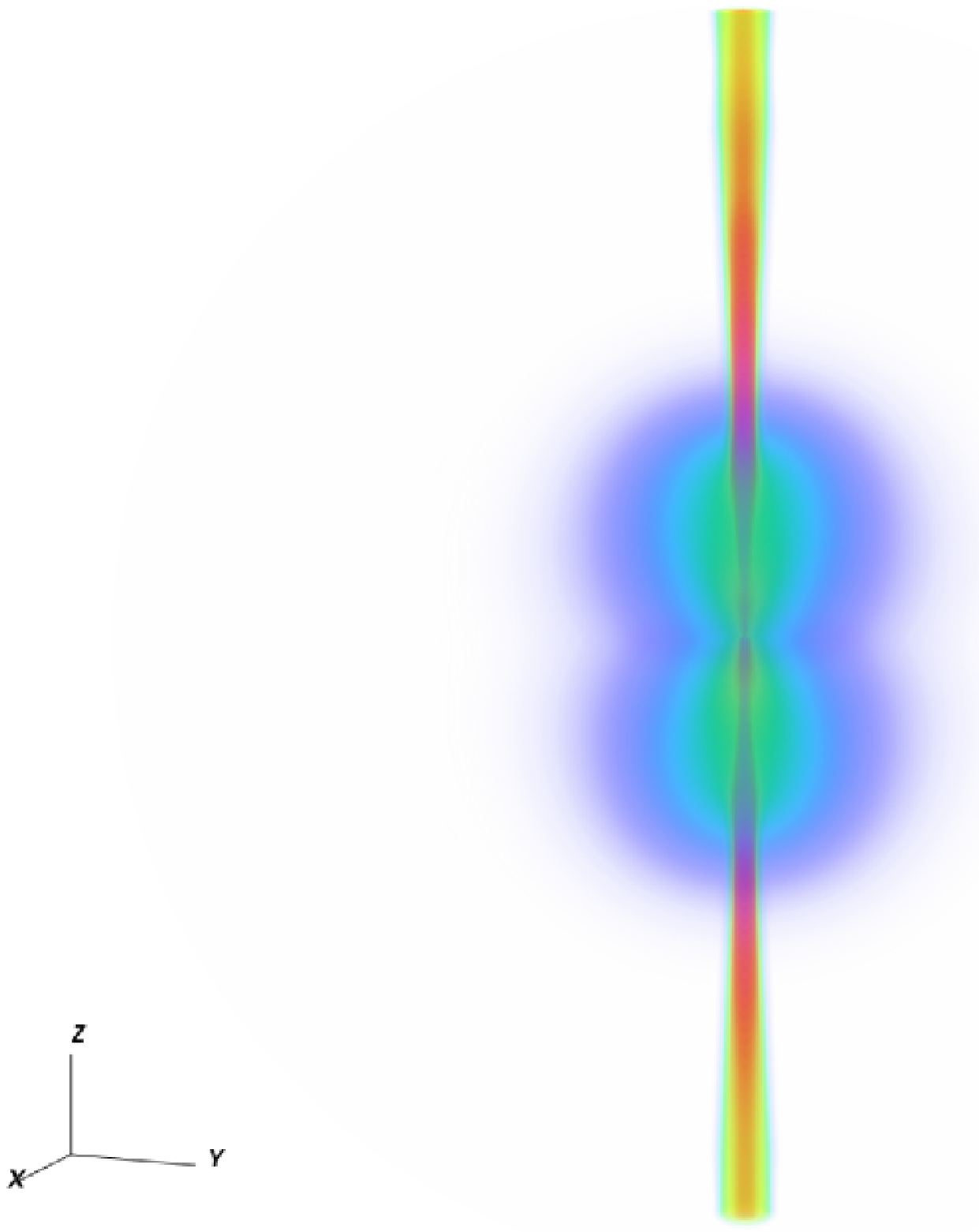}&
  \includegraphics[clip,height=0.35\textheight]{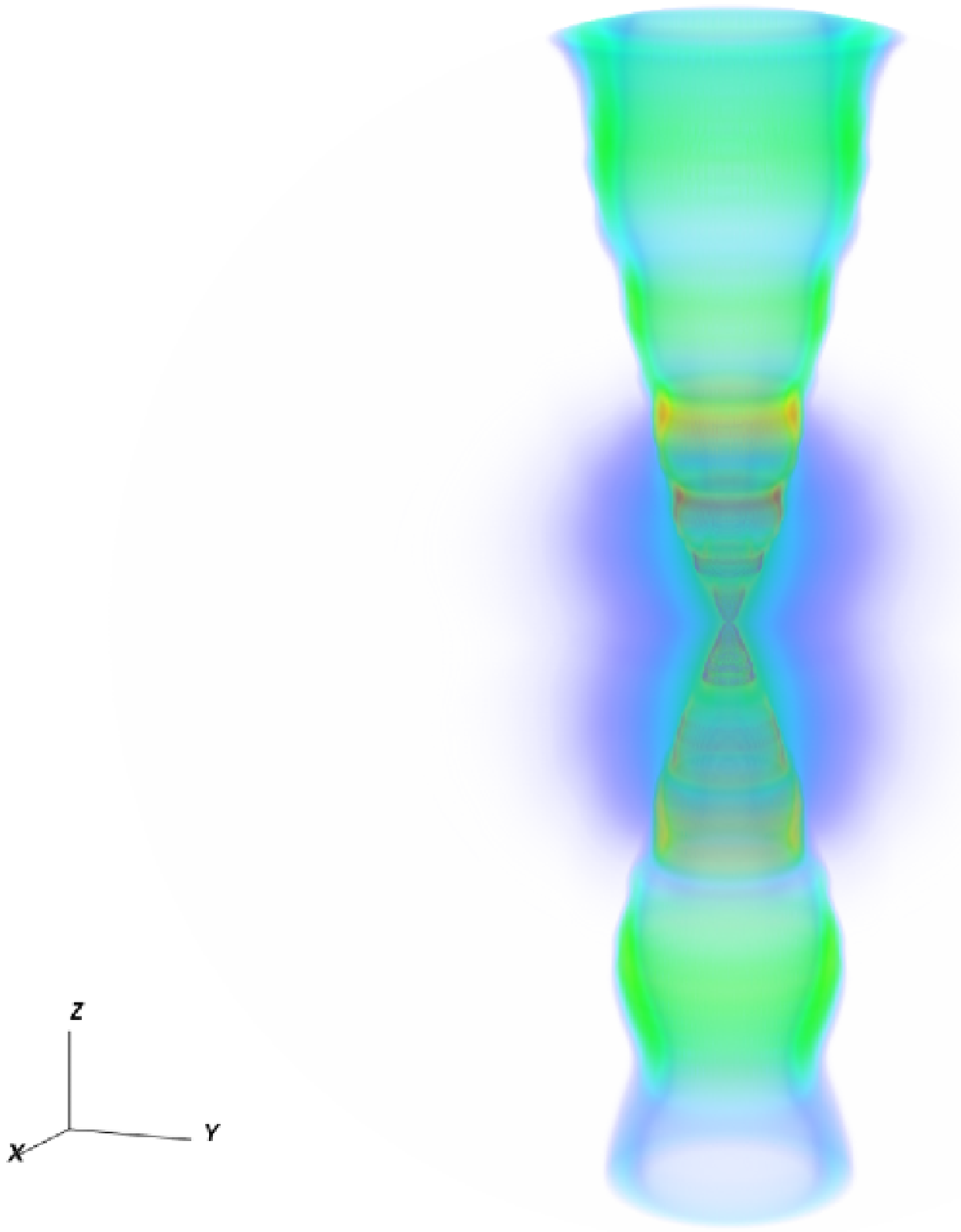}\tabularnewline
  \includegraphics[clip,height=0.35\textheight]{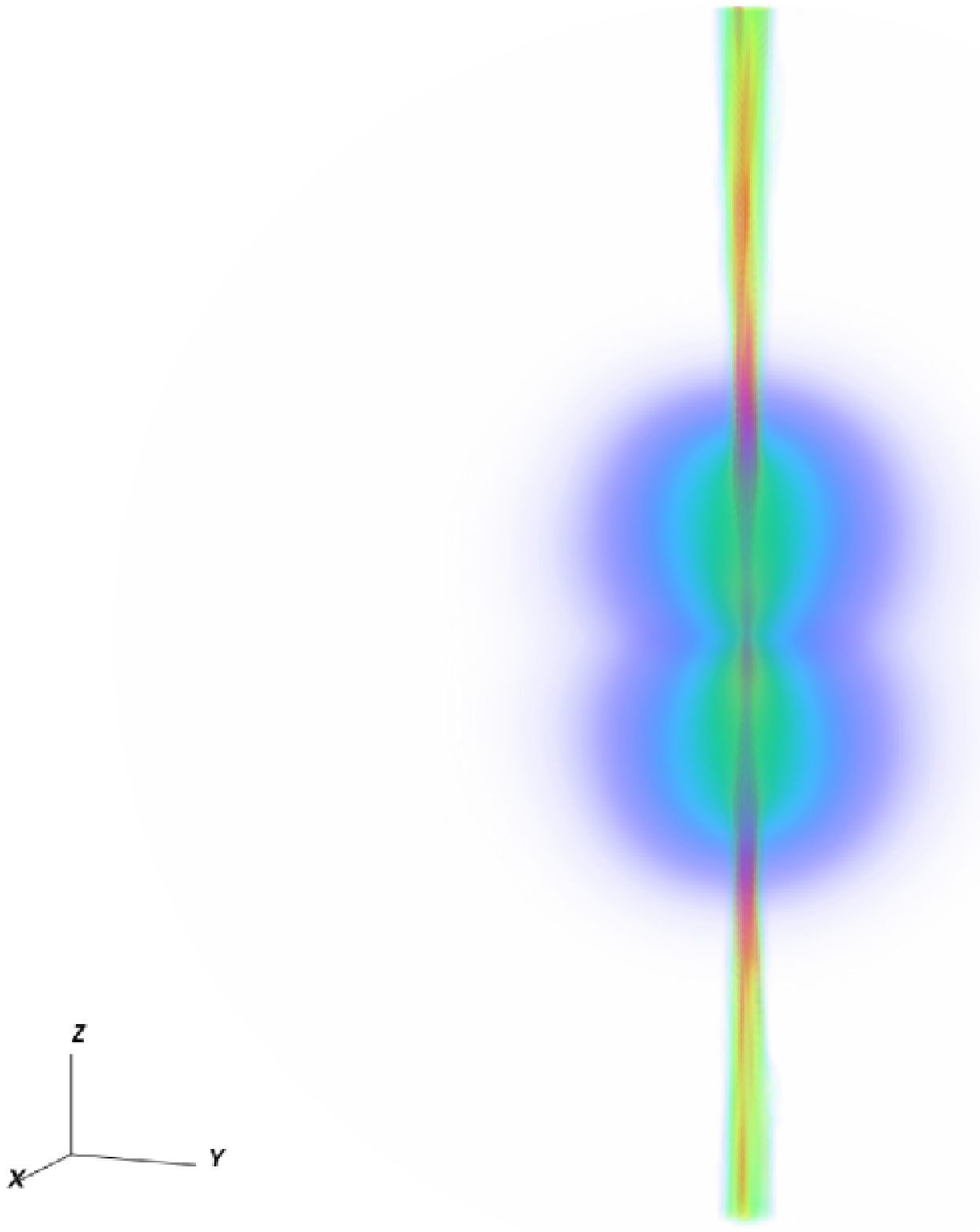}&
  \includegraphics[clip,height=0.35\textheight]{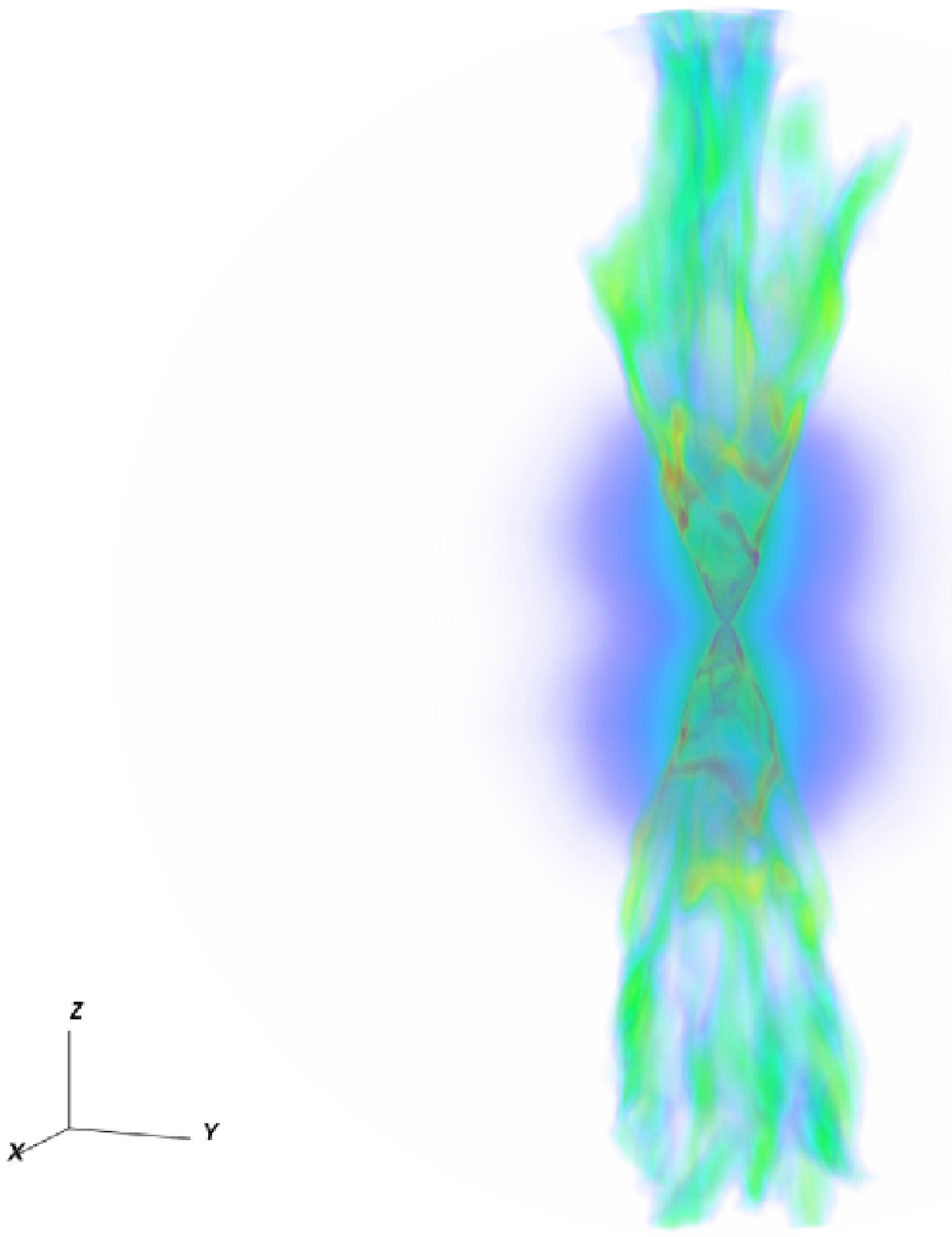}\tabularnewline
\end{tabular}

\end{center}

\caption{Density distributions from the 2-D (\emph{upper
panels}) and 3-D (\emph{lower panels}) models with (\emph{right panels})
and without (\emph{left panels}) gas rotation. The volume
rendering representation of the 3-D density distributions for Models
I, II, III and IV (cf.~Table~\ref{tab:Model-Summary}) are shown
in the \emph{upper-left, upper-right, lower-left} and \emph{lower-right}
panels, respectively. The 2-D models are assumed to be axisymmetric,
and the density values are extended around the symmetry axis to provide
full 3-D views. The length scale of each panel from the top to the
bottom is approximately 14~pc. }

\label{fig:Density-3d}

\end{figure*}


\subsection{Reference Values}

\label{sub:Reference-Values}


\placetable{tab:Model-Summary}

\begin{table*}

\caption{\label{tab:Model-Summary}Model Summary}

\begin{center}

\scriptsize

\begin{tabular}{clcccccccc}
\hline 
\hline&
$\left(n_{r},\, n_{\theta},\, n_{\phi}\right)$&
Rotation&
$\dot{M}_{\mathrm{in}}\left(r_{\mathrm{o}}\right)$&
$\dot{M}_{\mathrm{net}}\left(r_{\mathrm{i}}\right)$&
$\dot{M}_{\mathrm{out}}\left(r_{\mathrm{o}}\right)$&
&
$P_{k}\left(r_{\mathrm{o}}\right)$&
$P_{\mathrm{th}}\left(r_{o}\right)$&
\tabularnewline
Model&
$ $&
&
$\left(10^{25}\mathrm{\, g\, s^{-1}}\right)$&
$\left(10^{25}\,\mathrm{g\, s^{-1}}\right)$&
$\left(10^{25}\,\mathrm{g\, s^{-1}}\right)$&
&
$\left(10^{40}\,\mathrm{erg\, s^{-1}}\right)$&
$\left(10^{40}\,\mathrm{erg\, s^{-1}}\right)$&
\tabularnewline
\hline
I&
$128,128,1$&
no&
-10&
-1.8&
8.0&
&
94&
0.01&
\tabularnewline
II&
$128,128,1$&
yes&
-10&
-5.0&
5.8&
&
6.0&
0.21&
\tabularnewline
III&
$128,128,64$&
no&
-10&
-1.8&
8.0&
&
94&
0.01&
\tabularnewline
IV&
$128,128,64$&
yes&
-10&
-5.2&
5.3&
&
4.6&
0.27&
\tabularnewline
\hline
\end{tabular}

\normalsize

\end{center}
\end{table*}


The following parameters are common to all the models presented here,
and are exactly the same as in Paper~I and Paper~II.
We assume that the central BH is non-rotating and has mass $M_{\mathrm{BH}}=10^{8}\, M_{\odot}$.
The size of the disk inner radius is assumed to be $r_{*}=3r_{s}=8.8\times10^{13}\,\mathrm{cm}$
(c.f.~Sec.~\ref{sub:Model-Setup}). The mass accretion rate ($\dot{M}_{a}$)
of the central SMBH and the rest mass conversion efficiency ($\eta$)
are assumed to be $1\times10^{26}\,\mathrm{g\, s^{-1}}$ and $0.0833$,
respectively. With these parameters, the corresponding accretion luminosity
of the system is $L=7.5\times10^{45}\,\mathrm{erg\, s^{-1}=2\times10^{12}\,\Lsun}$.
Equivalently, the system has the Eddington number $\Gamma=0.6$ where
$\Gamma\equiv L/L_{\mathrm{Edd}}$ and $L_{\mathrm{Edd}}$ is the
Eddington luminosity of the Schwarzschild BH, i.e., $4\pi cGM_{\mathrm{BH}}/\sigma_{e}$.
The fractions of the luminosity in the UV ($f_{\mathrm{UV}}$) and
that in the X-ray ($f_{\mathrm{X}}$) are fixed at $0.95$ and $0.05$
respectively, as in Paper~I (their Run~C) and in Paper~II
(their Run~Cr). 

Important reference physical quantities relevant to our systems are
as follows. The Compton radius, $R_{C}\equiv GM_{\mathrm{BH}}\mu\, m_{p}/kT_{C}$,
is $8\times10^{18}\,\mathrm{cm}$ or equivalently $9\times10^{4}\, r_{*}$
where $T_{C}$, $\mu$ and $m_{p}$ are the Compton temperature, the
mean molecular weight of gas and the proton mass, respectively. We
assume that the gas temperature at infinity is
$T_{\infty}=T_{C}=2\times10^{7}\,\Kelvin$ and $\mu=1$. The
corresponding speed of sound at infinity is $c_{\infty}^{2}=(\gamma
kT_{C}/\mu m_{p})^{1/2}=4\times10^{7}\,\mathrm{cm\, s^{-1}}$.   The
corresponding Bondi radius \citep{Bondi:1952} is 
$R_{B}=GM_{\mathrm{BH}}/c_{\infty}^{2}=4.8\times10^{18}\,\mathrm{cm}$
while its relation to the Compton radius is $R_{B}=\gamma^{-1}R_{C}$.
The Bondi accretion rate (for the isothermal flow) is $\dot{M}_{B}=3.3\times10^{25}\,\mathrm{g\, s^{-1}}=0.52\,\MsunPerYear$.
The corresponding free-fall time ($t_{\mathrm{ff}}$) of gas from
the Bondi radius to the inner boundary is $2.1\times10^{11}\,\mathrm{sec}=7.0\times10^{3}\,\mathrm{yr}$.
The escape velocity from the inner most radius ($r_{\mathrm{i}}=500\, r_{*}$)
of the computational domain is about $7.7\times10^{4}\,\,\mathrm{km\, s^{-1}}$.


\begin{figure*}
\begin{center}

\begin{tabular}{cc}

  \includegraphics[clip,height=0.3\textheight]{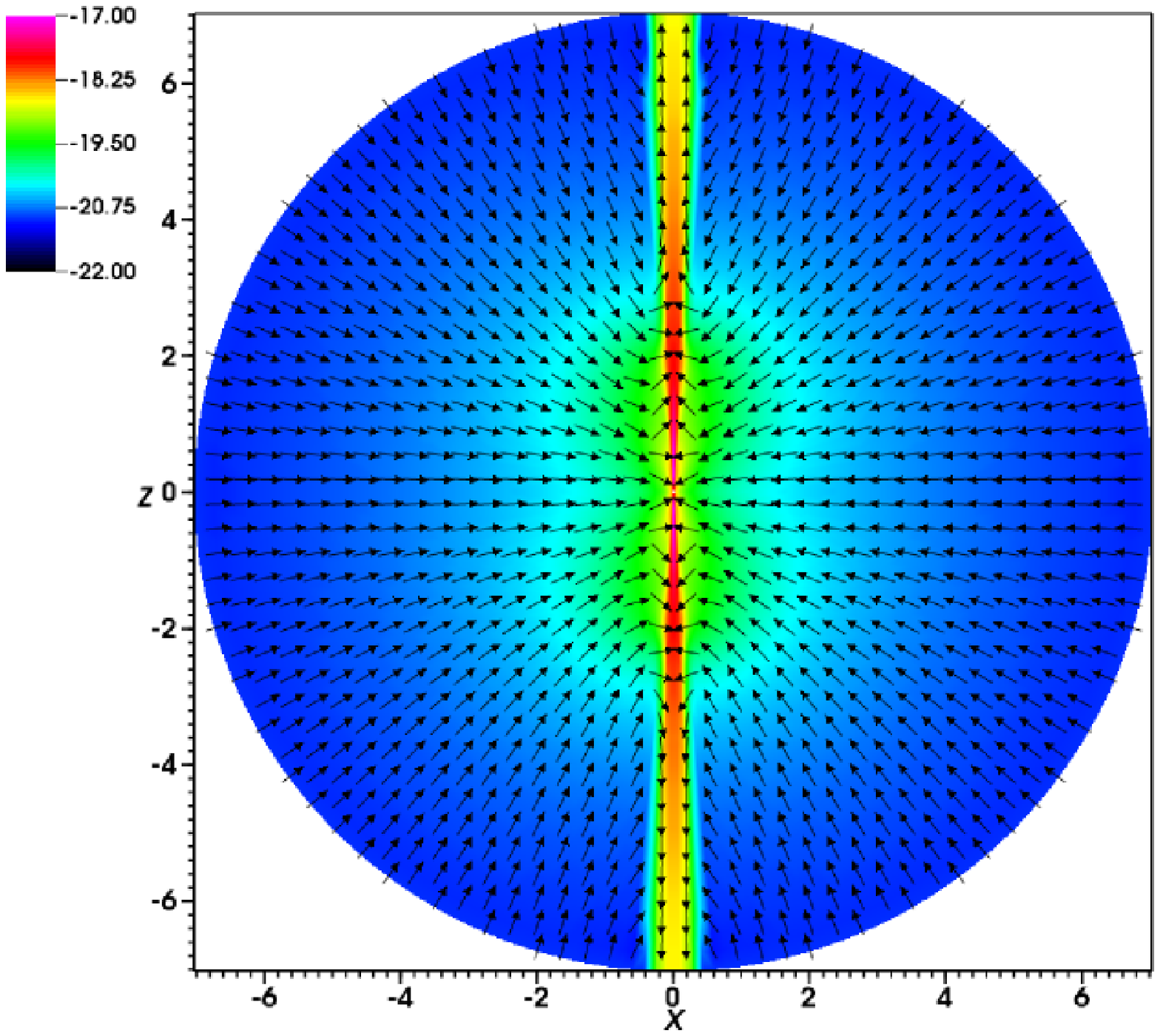}&
  \includegraphics[clip,height=0.3\textheight]{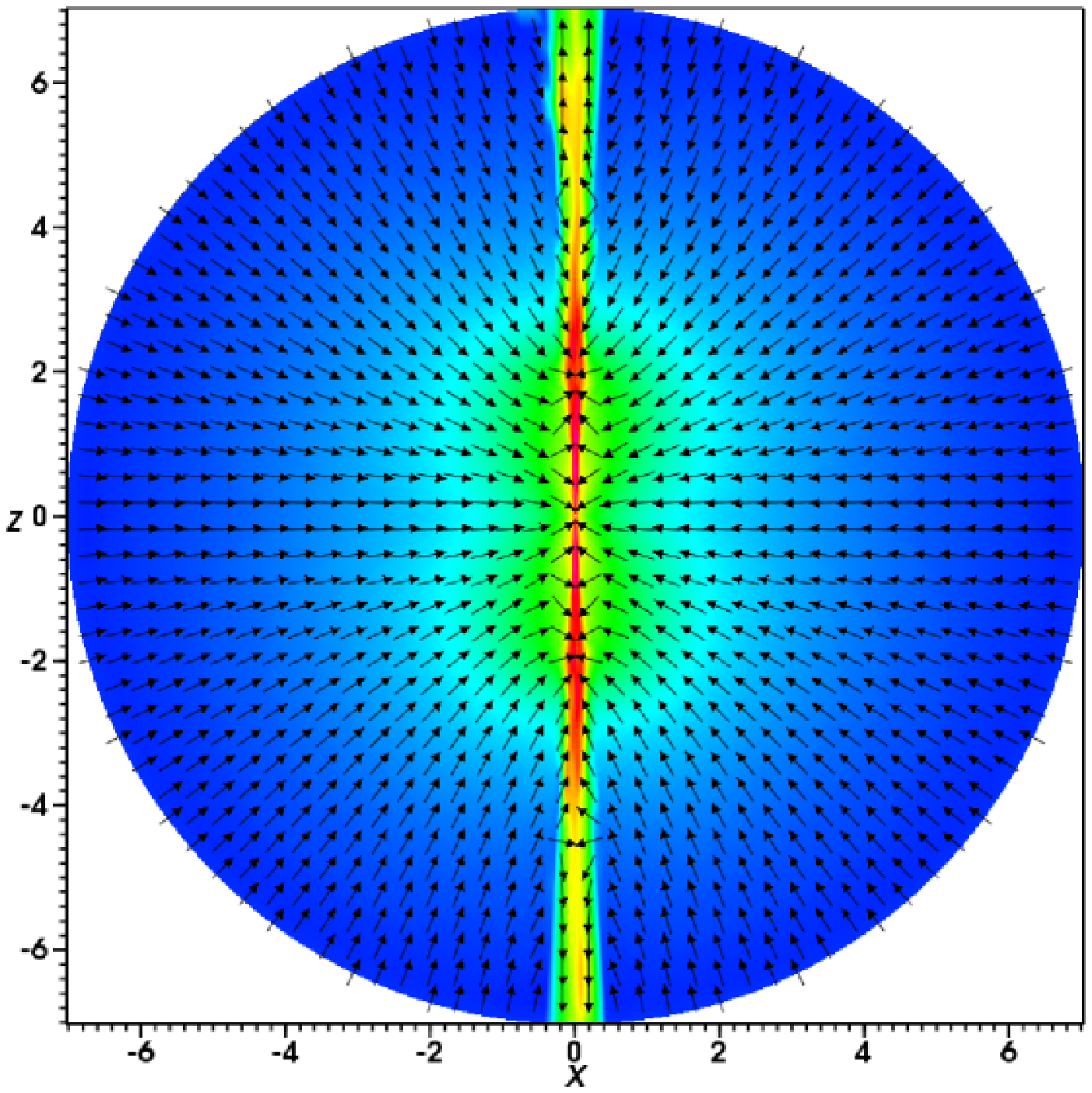}\tabularnewline
  \includegraphics[clip,height=0.3\textheight]{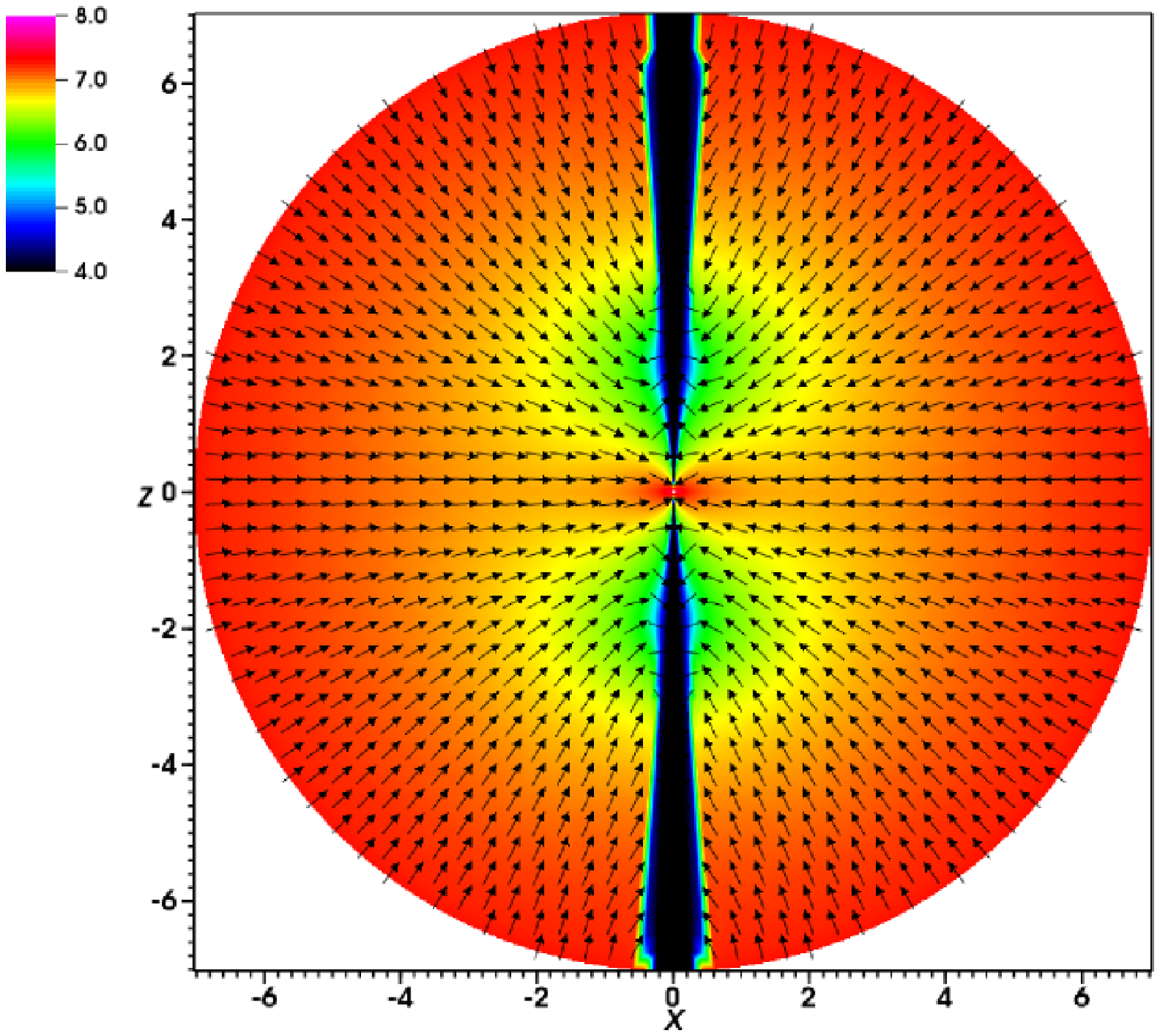}&
  \includegraphics[clip,height=0.3\textheight]{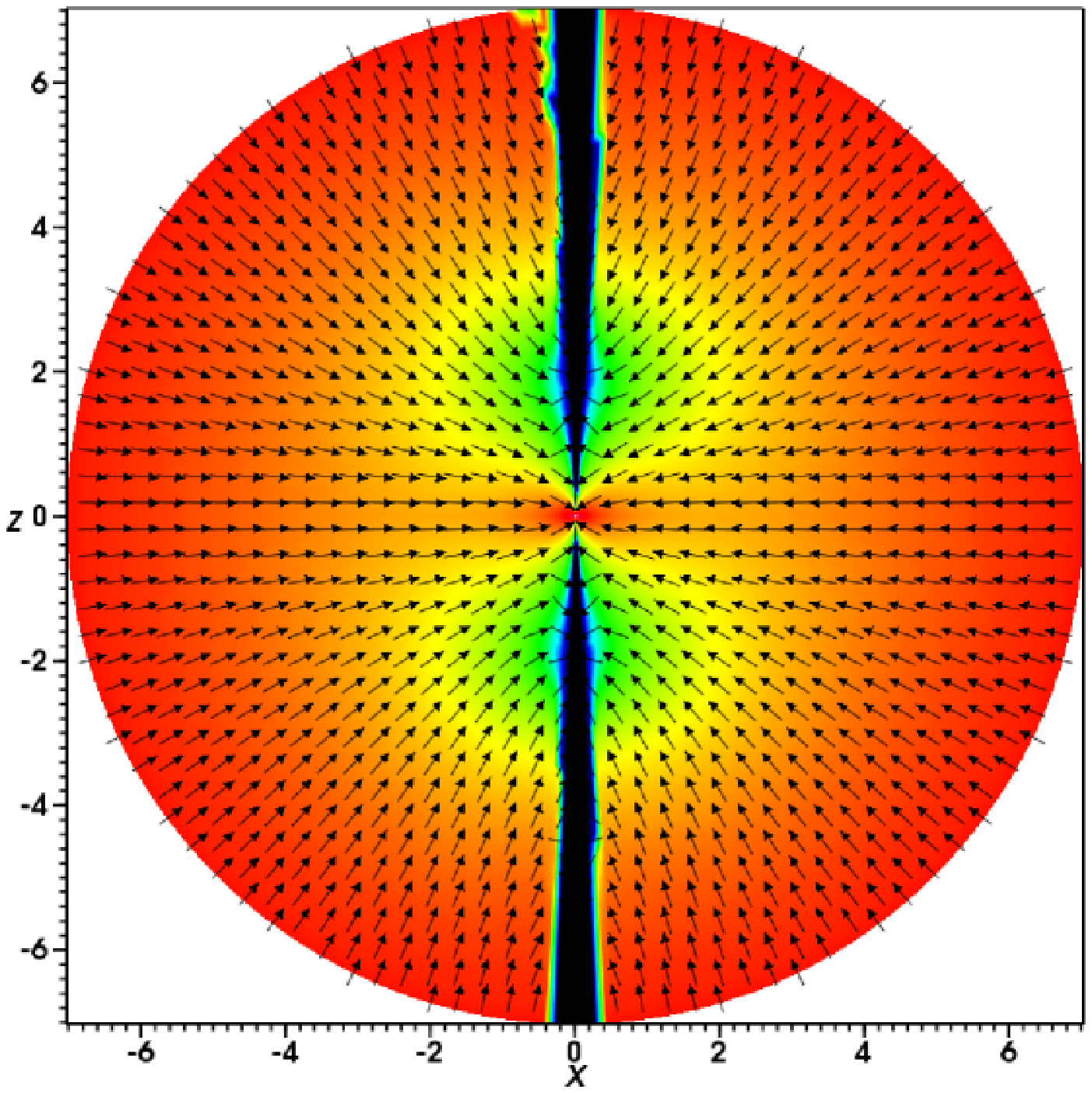}\tabularnewline

\end{tabular} 

\end{center}

\caption{Comparison of the results from the non-rotating models:
  Models~I (\emph{left panels}) and III (\emph{right panels}). The
  density (\emph{upper panels}) and temperature maps (\emph{lower
    panels}) in logarithmic scale \textcolor{black}{(in cgs units)} are
  overplotted by the directions of the poloidal velocity as black
  arrows.  The length scales are in pc.  Overall structures of the
  density and temperature are very similar to each other. Both models
  show rather narrow outflows in the polar directions, and the inflows
  in the equatorial regions. The 3-D model (Model~III) shows a small
  but noticeable amount of non-axisymmetric density and temperature
  distributions in the narrow cones of the outflowing regions in the
  polar directions. The opening angles of the outflows in both cases
  are $\sim5^{\circ}$. }

\label{fig:Density-Temp-Map-I-III}

\end{figure*}



\begin{figure*}
\begin{center}

\begin{tabular}{cc}

  \includegraphics[clip,height=0.3\textheight]{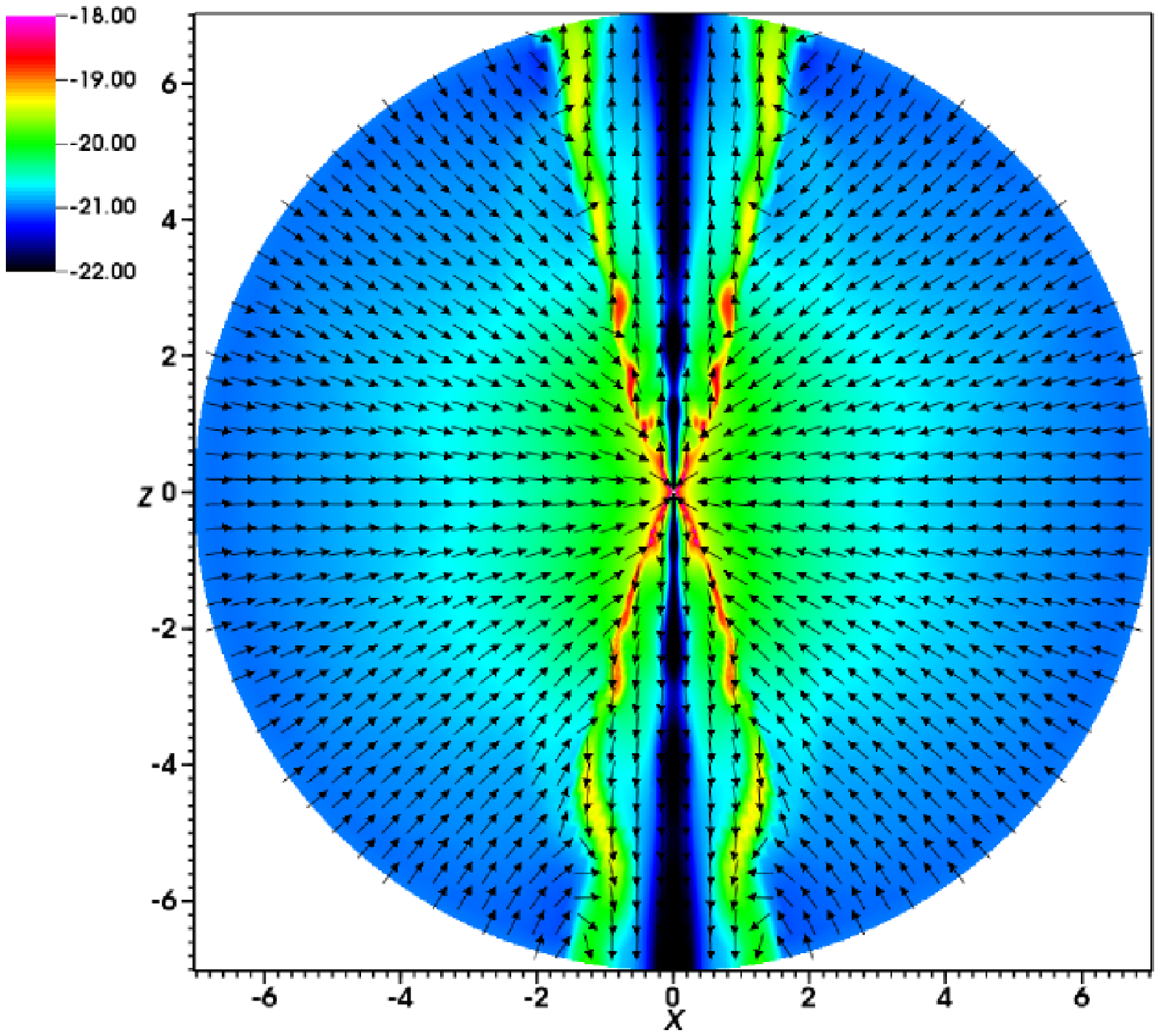}&
  \includegraphics[clip,height=0.3\textheight]{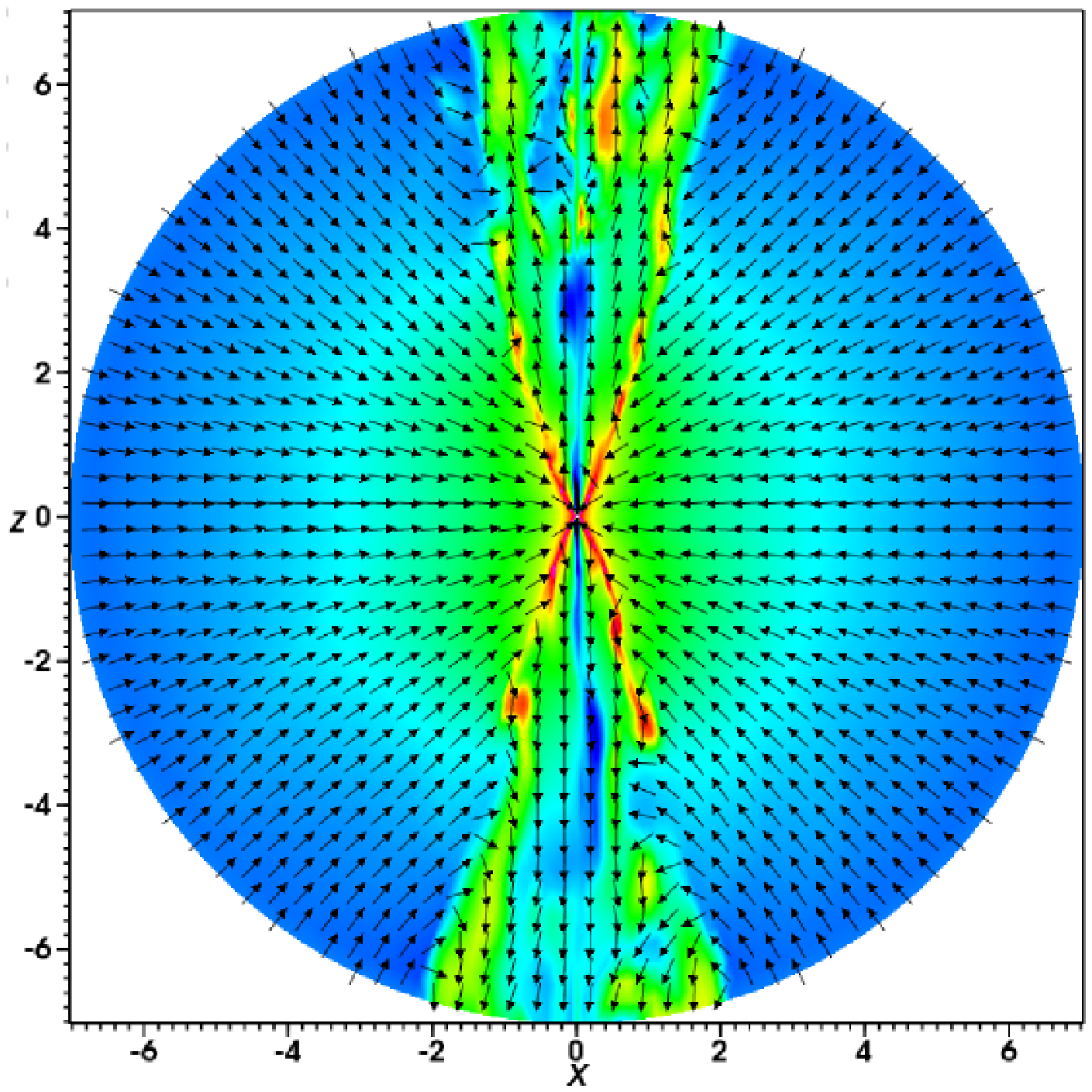}\tabularnewline
  \includegraphics[clip,height=0.3\textheight]{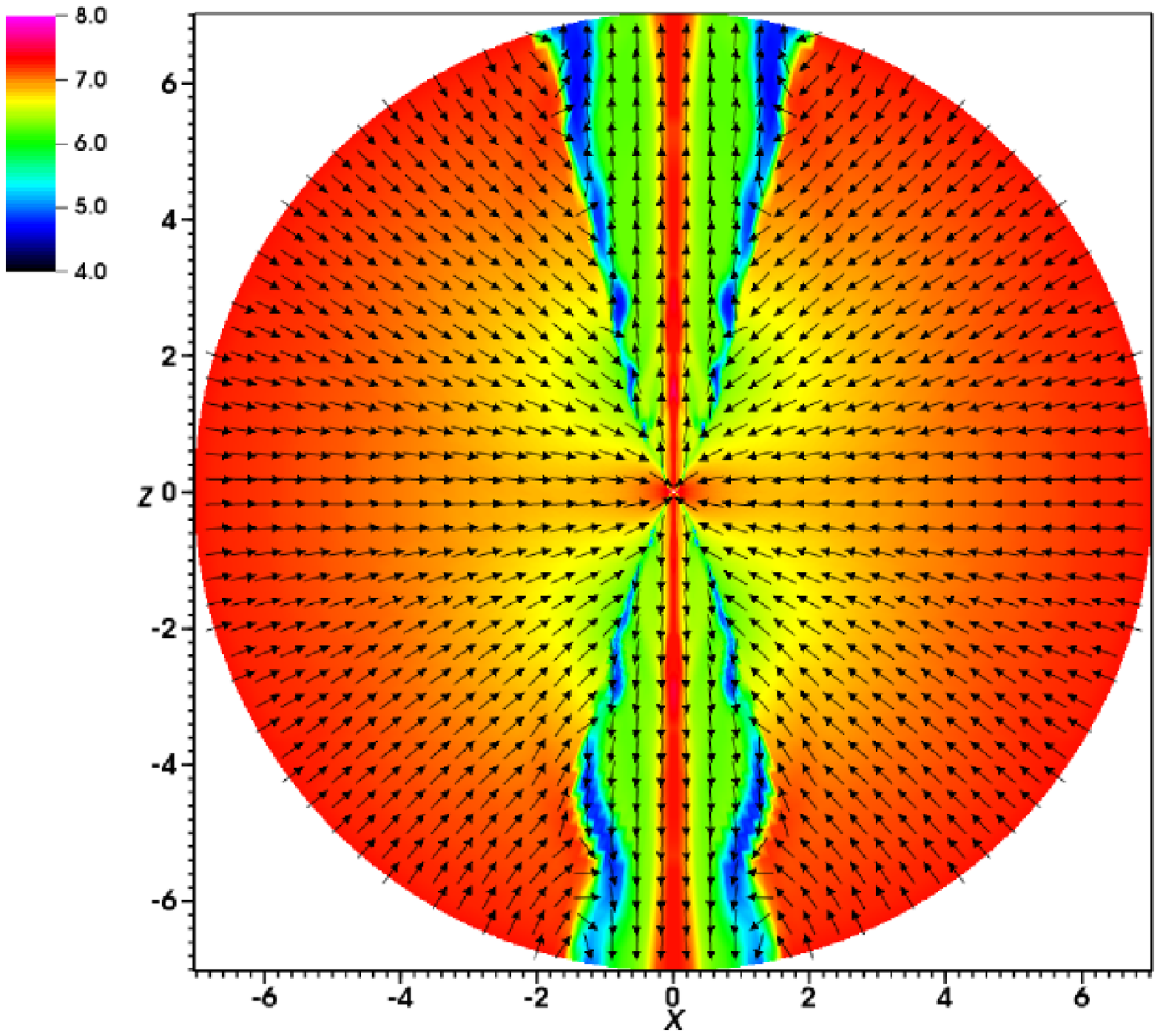}&
  \includegraphics[clip,height=0.3\textheight]{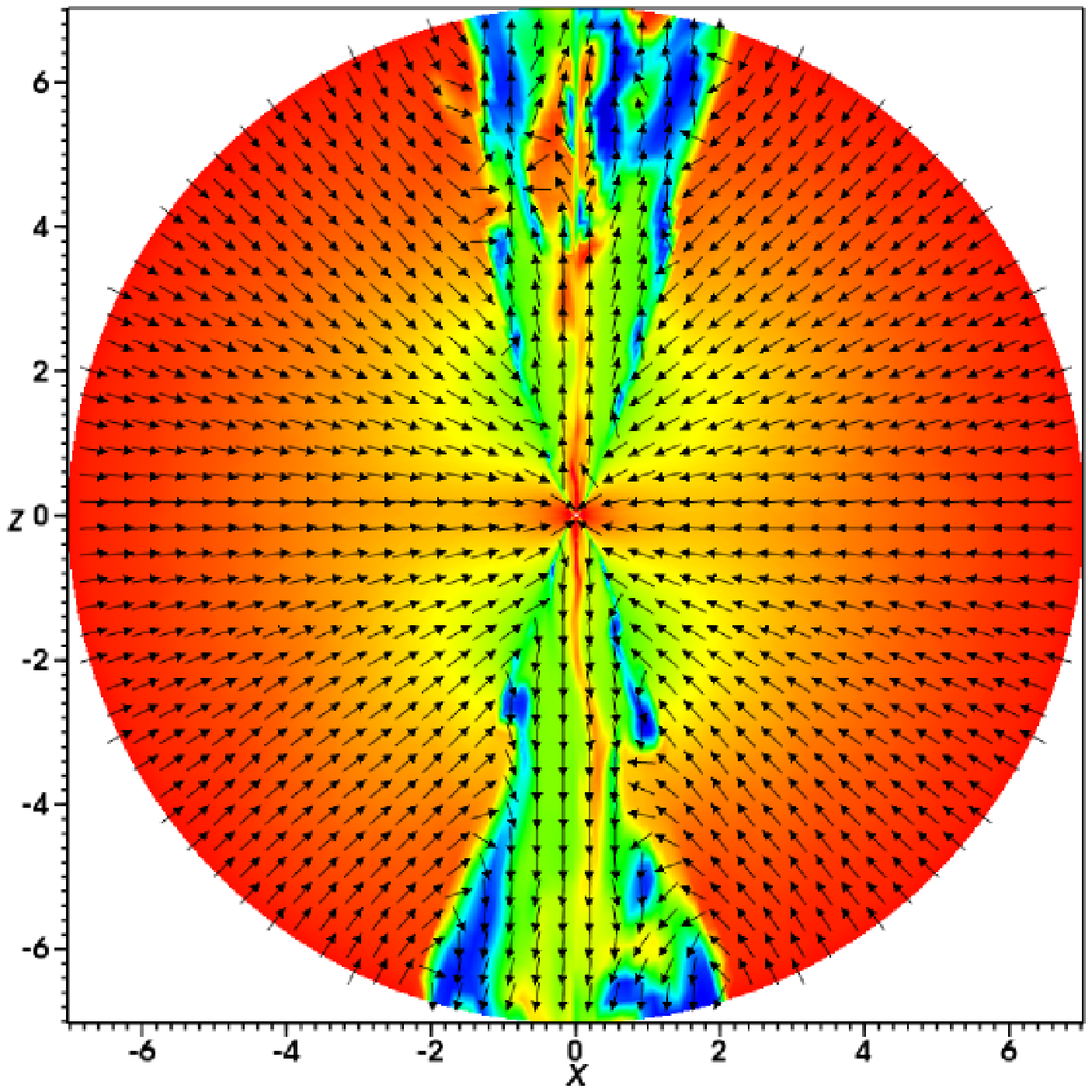}\tabularnewline

\end{tabular}

\end{center}

\caption{As in \textcolor{black}{Fig.~\ref{fig:Density-Temp-Map-I-III}},
but for Models~II (\emph{left panels}) and IV (\emph{right panels})
in which the rotation of gas is included. Compared to the non-rotating
models (Fig.~\ref{fig:Density-Temp-Map-I-III}), the outflows seen
here are less collimated, and the higher density clumpy structures
with lower temperatures moves outwards along (and near) conic surfaces.
The non-axisymmetric nature of the flows for the 3-D model (Model~IV)
is clearly seen. The opening angles of the outflows in both cases are
$\sim30^{\circ}$.}

\label{fig:Density-Temp-Map-II-IV} 

\end{figure*}

\subsection{Density, Temperature and Velocity Structures}

\label{sub:density-temp-maps}

The 3-D representations of the density (as volume rendering images)
of the models are shown in Figure~\ref{fig:Density-3d}. For the
2-D models, the density is extended around the $z$-axis using the
axisymmetry, to give 3-D views. The corresponding density and temperature
maps along with the directions of the poloidal velocity of the flows
on the $z$--$x$ plane are given in Figures~\ref{fig:Density-Temp-Map-I-III}
and \ref{fig:Density-Temp-Map-II-IV}. 

For non-rotating gas cases (Models~I and III), the
outflow occurs in 
very narrow cones in the polar directions 
\textcolor{black}{(Figs.~\ref{fig:Density-3d} and
  \ref{fig:Density-Temp-Map-I-III})}. 
The opening angles of the
outflows in both models are about $5^{\circ}$. The figures show that
overall density structures of Models~I and III are very similar to each
other. Small but noticeable differences can be seen in the density
structure in the narrow outflow regions. While the flow in Model~I
(2-D) is very smooth (steady), that of Model~III (3-D) shows a hint of
unsteadiness as \textcolor{black}{indicated by} the non-monotonic change of the density
along the pole directions (unlike that of Model~I). The increase of
unsteadiness \textcolor{black}{in the outflows of the 3-D model} can 
be also seen in the variability of the 
mass outflow flux which we will discuss later in
\S~\ref{sub:Mass-and-Energy}.  Model~III also shows a sign of
non-axisymmetric flow although the degree of non-axisymmetry is rather
small [$\sim 38$~\% variation of $\rho$ around the rotation axis for
  $r=10^{4}\,r_{*}$ and $\theta=5^{\circ}$
  (cf.~\S~\ref{sub:Non-axisymetric-nature})]. This can be clearly seen
in the density
\textcolor{black}{(Fig.~\ref{fig:Density-Temp-Map-I-III})} of the narrow
cones near the outer boundary where the 
density across a horizontal line is not symmetric with respect to the
poles (the $z$-axis).  In spite of the small non-axisymmetry and
variability of the internal structure of the narrow outflow cones, we
find the overall structure or the integrity of the narrow outflow
cones are intact, i.e., we find no wiggling of the cones themselves.

The gas rotation dramatically changes the morphology of the outflows.
The centrifugal force due to gas rotation evidently pushes outflows
away from the polar axis, and forms much wider outflows (less collimated),
as seen in Figures~\ref{fig:Density-3d} and \ref{fig:Density-Temp-Map-II-IV}.
The opening angles of the outflows in both models are approximately $30^{\circ}$.
While the density is relatively high in the polar directions for the
non-rotating models (Models~I and III), it is relatively low for
the rotating models (Models~II and IV). The higher density regions
(for the rotating cases) occur on and near the conic surfaces formed
both above and below the equatorial planes. Similarly, the temperature
along the poles is relatively \textcolor{black}{low} for the non-rotating cases, but
it is relatively \textcolor{black}{high} for the rotating cases, especially in 2-D cases.
Essentially the same differences between the models with and without
 gas rotation are found by of Paper~II, cf., their run~C
and Cr. 

As also observed in the model of Paper~II, we find the
outflows in the rotating cases tend to be fragmented into smaller
pieces which have relatively high density and relatively low temperature
(see Fig.~\ref{fig:Density-Temp-Map-II-IV}). We find that these cold
{}``cloud-like'' features are formed around
$z\approx1.5\times10^{4}\, r_{*}$, and they flow outward along the
outflow conic surface. We also find that the clouds (adiabatically)
cool and expand as they move outward (see
\S~\ref{sub:Photoionization-paramters}). Fig.~4 of Paper~II, 
showing a time-sequence of density maps, demonstrates the motion
of the cold outflow.  The fragmentation of the outflow in the models with
gas rotation (Model~II and IV) is caused by a rapid radiative cooling
of a high density gas which is formed at location where the inflow
turns into the outflow, and the geometry of the outflow (the curved
shape) which allows for a quite direct exposure to the strong X-ray
from the central source. Readers are referred to Paper~II
for a more detailed explanation for the cause of fragmentation. 

This cloud-like feature seen in the 2-D maps, of course, will look
like rings if the density is rotated around the symmetry axis, as
seen in the 3-D representation of the 2-D model with gas rotation
(Model~II in Fig.~\ref{fig:Density-3d}). In the 3-D model with
rotation (Model~IV in Fig.~\ref{fig:Density-3d}), we find that
this ring structure is not stable. The ring tends to be deformed and
breaks connections, due to shear and thermal instabilities. 
The parts of the broken ring structure also have relatively high
density and low temperatures. They also resembles rather elongated
cold cloud-like structures. Although the overall density and temperature
structure of the flows in 2-D and 3-D for rotating cases are very
similar to each other, the outflows occur in much less organized manner
in the 3-D model.

\subsection{Mass and Energy Flux}

\label{sub:Mass-and-Energy}

To examine the characteristics of the flows in the models more qualitatively,
we compute the mass fluxes as a function of radius. For the 3-D models,
the net mass flux ($\dot{M}_{\mathrm{net}}$), the inflow mass flux
($\dot{M}_{\mathrm{in}}$) and the outflow mass flux ($\dot{M}_{\mathrm{out}}$)
are computed by following Paper~I (see also \citealt{Kurosawa:2008}),
\begin{align}
  \dot{M}\left(r\right) &=\oint_{s}\rho\,\boldsymbol{v\,\cdot}d\boldsymbol{a}
  \label{eq:mdot}\\
                        & =r^{2}\oint_{4\pi}\rho v_{r}\, d\Omega,
  \label{eq:mdot2}
\end{align}
where $v_{r}$ is the radial component of velocity $\boldsymbol{v}$.
The net mass flux is obtained in the equation above if all $v_{r}$
are included. Similarly, the inflow mass flux and the outflow flux
are obtained if only the points with $v_{r}<0$ and with $v_{r}>0$
are included, respectively, in the integration. The surface element
and the solid angle element are $d\boldsymbol{a}=\boldsymbol{\hat{r}}\, r^{2}\sin\theta\, d\theta\, d\phi$
and $d\Omega=\sin\theta\, d\theta\, d\phi$. We further define the
\emph{outflow} power in the form of kinetic energy ($P_{k}$) and
that in the thermal energy ($P_{\mathrm{th}}$) as functions of radius,
i.e., 
\begin{align}
  P_{k}\left(r\right) & =r^{2}\oint_{4\pi}\rho v_{r}^{3}\, d\Omega
  \label{eq:power-kinetic}
\end{align}
and 
\begin{align}
  P_{\mathrm{th}}\left(r\right) & =r^{2}\oint_{4\pi}e\,v_{r}\,d\Omega\,.
  \label{eq:power-thermal}
\end{align}
where $v_{r}>0$. For the 2-D models, the integrations are \textcolor{black}{performed} by
assuming the axi-symmetry. 

\begin{figure*}
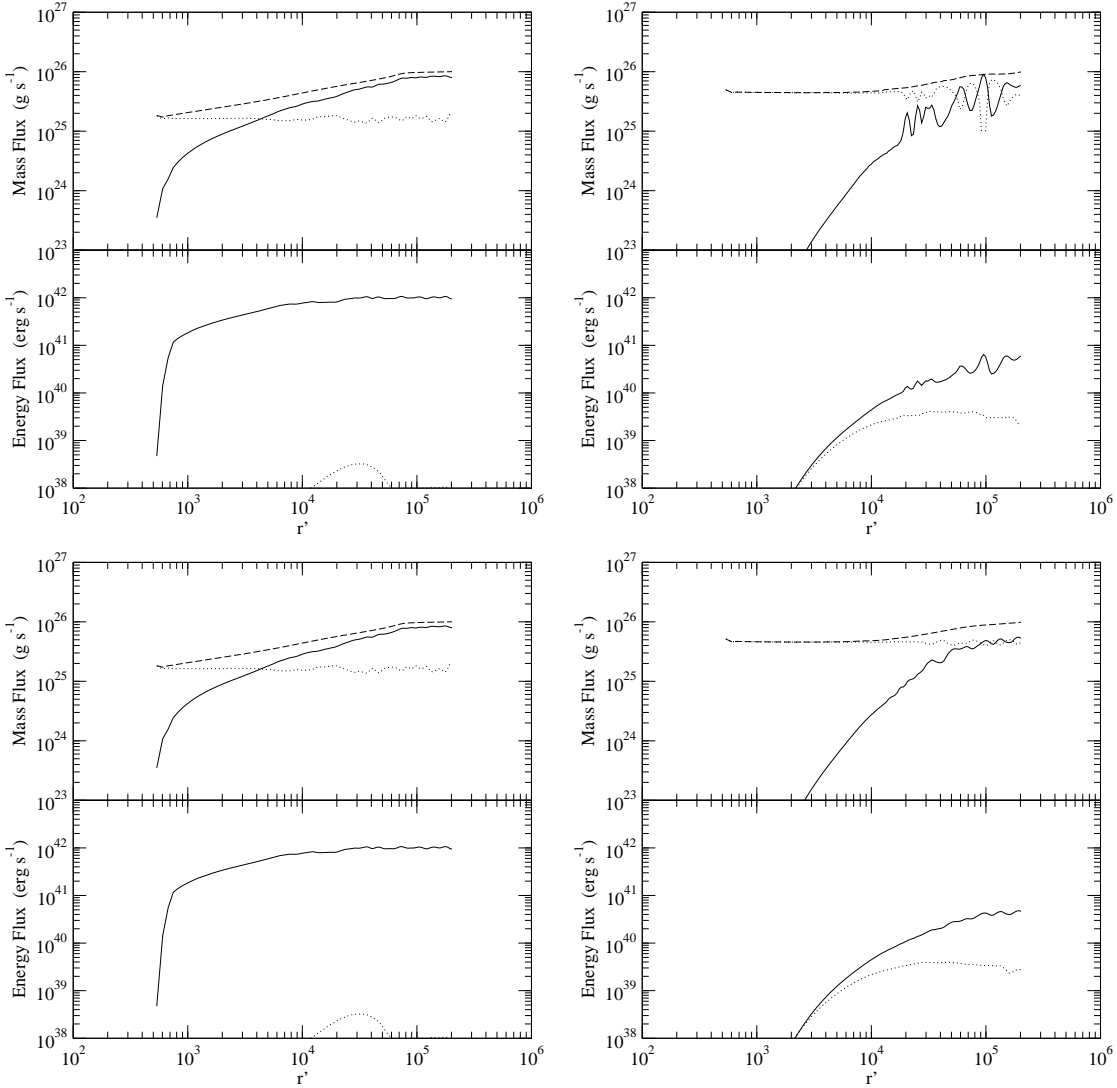

\begin{center}

\begin{tabular}{cc}
  \includegraphics[clip,width=0.4\textwidth]{f05a.eps}&
  \includegraphics[clip,width=0.4\textwidth]{f05b.eps}\vspace{0.1cm}\tabularnewline
  \includegraphics[clip,width=0.4\textwidth]{f05c.eps}&
  \includegraphics[clip,width=0.4\textwidth]{f05d.eps}\tabularnewline
\end{tabular}

\end{center}

\caption{Comparison of the mass and energy fluxes
as a function of radius for Models~I (\emph{upper left}), II (\emph{upper
right}), III (\emph{lower left}), IV (\emph{lower right}). Each panel
is subdivided into two parts: top (mass flux) and bottom (energy flux).
In the mass flux plots, the inflow (\emph{dashed line}; $\dot{M}_{\mathrm{in}}$),
outflow (\emph{solid line}; $\dot{M}_{\mathrm{o}}$) and net (\emph{dotted
lines}; $\dot{M}_{\mathrm{net}}$) mass fluxes, as defined in eq.~(\ref{eq:mdot2}),
are separately plotted, as a function of radius. The absolute values
of $\dot{M}_{\mathrm{in}}$ and $\dot{M}_{\mathrm{net}}$ are plotted
here since they are negative at all radii. The length scale is in
units of the inner disk radius ($r'=r/r_{*}$). In the energy flux
plots, the kinetic energy (\emph{solid line}) and the thermal energy
(\emph{dotted line}) fluxes, defined as eqs.~(\ref{eq:power-kinetic})
and (\ref{eq:power-thermal}), are shown. Note that the time slices
of the model simulations used here to computed the fluxes are same
as those in Figures~\ref{fig:Density-3d}, \ref{fig:Density-Temp-Map-I-III}
and \ref{fig:Density-Temp-Map-II-IV}.}

\label{fig:mass-energy-flux}

\end{figure*}

\begin{figure}
\begin{center}

\includegraphics[clip,width=0.48\textwidth]{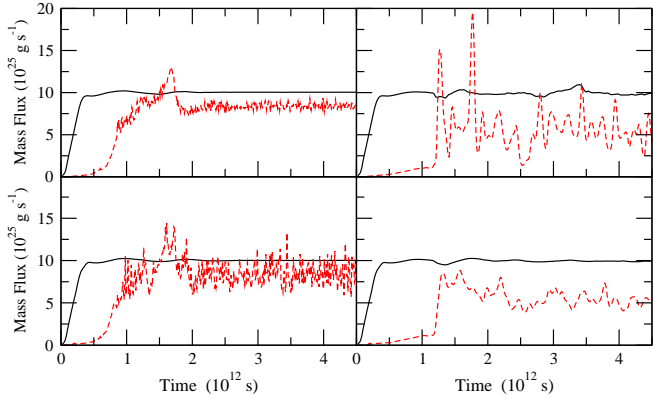}

\end{center}

\caption{The mass flow rates across the outer boundary
(cf.~eq.~{[}\ref{eq:mdot2}]) as a function of time for Models~I
(\emph{upper left}), II (\emph{upper right}), III (\emph{lower left}),
and IV (\emph{lower right)}. Each panel shows the mass inflow rate
at the outer boundary (\emph{solid line}), and the mass outflow rate
at the outer boundary (\emph{dashed line}). The mass-inflow rates
of the four models are almost constant for time~$>10^{12}\,\mathrm{s}$,
and their values are almost identical to each other ($\sim10^{26}\,\mathrm{g\, s^{-1}}$).
On the other hand, the mass outflow rates show variability. The amplitudes
of the variability are relatively larger in the 3-D model compared to
those in the 2-D model for the non-rotating cases while the opposite is seen for the rotating cases.
The average mass outflow rates for the non-rotating cases ($\sim8\times10^{25}\,\mathrm{g\, s^{-1}}$)
are slightly larger than that of the rotating cases ($\sim5\times10^{25}\,\mathrm{g\, s^{-1}}$)
at a later time in the simulation (i.e., time~$>3\times10^{12}\,\mathrm{s}$).
 }

\label{fig:mdot-evolution}

\end{figure}

The resulting mass fluxes and the outflow powers of the models are
summarized in Figure~\ref{fig:mass-energy-flux}. In all cases, the
mass inflow flux ($\dot{M}_{\mathrm{in}}$) exceeds the mass outflow
rate ($\dot{M}_{\mathrm{out}}$) at all radii, except for the one
point at $r'(=r/r_{*})\sim10^{5}$ for Model~II. For Models~I, III and IV,
the net mass fluxes ($\dot{M}_{\mathrm{net}}$) are almost constant
at all radii, indicating that the flows in these models are almost
steady. A relatively steady nature of the flows in these models can
be also seen in the time evolution of the mass inflow and outflow
fluxes at the outer boundary, i.e., $\dot{M}_{\mathrm{in}}\left(r_{\mathrm{o}}\right)$
and $\dot{M}_{\mathrm{out}}\left(r_{\mathrm{o}}\right)$, as shown
in Figure~\ref{fig:mdot-evolution}. 

We find that the radial dependencies of $\dot{M}_{\mathrm{in}}$,
$\dot{M}_{\mathrm{out}}$ and $\dot{M}_{\mathrm{net}}$ (Fig.~\ref{fig:mass-energy-flux})
of Model~III (3-D) are also almost identical to those of Model~I
(2-D). In \S~\ref{sub:density-temp-maps}, we found a hint
of non-uniform density variation along the narrow outflow cones in
the polar directions for the 3-D non-rotating case (Model~\textcolor{black}{III}). As
one can see from Figure~\ref{fig:mdot-evolution}, the time variability
in $\dot{M}_{\mathrm{out}}\left(r_{\mathrm{o}}\right)$ for Model~III
is slightly higher than that of the 2-D mode (Model~I). However,
we find that the time averaged values (between $t=3\times10^{12}$
and $4\times10^{12}$~s) of $\dot{M}_{\mathrm{out}}\left(r_{\mathrm{o}}\right)$
for the non-rotating models (Models~I and \textcolor{black}{III}) are almost identical
to each other. 

On the other hand, the 2-D rotating case (Model~II) in Figure~\ref{fig:mass-energy-flux}
shows a non-uniform distribution of $\dot{M}_{\mathrm{net}}$ for
$r'\gtrsim 10^{4}$. This is caused by the non-uniform distribution of
the outflow mass flux $\dot{M}_{\mathrm{out}}$ in $r'$, but not
by that the inflow mass flux $\dot{M}_{\mathrm{in}}$ which has a
smooth distribution across all radii. The non-uniform distribution
of $\dot{M}_{\mathrm{out}}$ (bumps) is caused by the presence of
the cold cloud-like (Fig.~\ref{fig:Density-Temp-Map-II-IV}) or 
ring-like (\textcolor{black}{Fig.~\ref{fig:Density-3d}}) structures in the outflow.
This also leads to a relatively large time variability in the outflow
mass flux at the outer boundary for Model~II, as shown in Figure~\ref{fig:mdot-evolution}.
Interestingly, the bumps in $\dot{M}_{\mathrm{out}}$ seen in Model~II
(Fig\@.~\ref{fig:mass-energy-flux}) are much less prominent in
the 3-D equivalent of this model (Model~IV). As mentioned before,
the very organized ring-like structures seen in the outflows of the
rotating 2-D model (Model~II) tend to be stretched and fragmented
in both radial and azimuthal directions (cf., Fig.~\ref{fig:Density-3d}).
The outflow becomes much less organized. This results in the smoothing
of the bumps on the $\dot{M}_{\mathrm{out}}$ curve in Figure~\ref{fig:mass-energy-flux}
for the 3-D model (Model~IV). This also causes the decrease in the
degree of the time-variability in the mass outflow flux at the outer
boundary, $\dot{M}_{\mathrm{out}}\left(r_{\mathrm{o}}\right)$, as
seen in Figure~\ref{fig:mdot-evolution}. Except for the bumps, overall
behaviors of the mass flux curves (as a function of radius) of Model~IV
are very similar to those of the 2-D model, Model~II. 
This shows that dimensionality does not change the gross properties of
radiation-driven winds, and is consistent with the results of 
\citet{Proga:1999} who studied radiation-driven winds in 1-D and
2-D.

The net mass fluxes at the inner boundary $\dot{M}_{\mathrm{net}}\left(r_{\mathrm{i}}\right)$
are $-1.8$, $-5.0$, $-1.8$ and $-5.2\times10^{25}\,\mathrm{g\, s^{-1}}$
(or equivalently $-0.30$, $-0.83$, $-0.30$ and $-0.87$~$\MsunPerYear$)
for Models~I, II, III and IV respectively
(Tab.~\ref{tab:Model-Summary}).  This indicates that the net mass flux
inward (negative signs indicate inflow) 
significantly increases when the gas is rotating (Models II and IV).
We also find that the inflow mass fluxes at the outer boundary
$\dot{M}_{\mathrm{in}}\left(r_{\mathrm{o}}\right)$ 
are same for all models ($-10\times10^{25}\,\mathrm{g\, s^{-1}}$), but
the outflow fluxes at the outer boundary
$\dot{M}_{\mathrm{out}}\left(r_{\mathrm{o}}\right)$ decreases when the
gas rotates (Tab.~\ref{tab:Model-Summary}).  The ratios of the total
mass outflow flux to the total mass inflow flux at the outer boundary
($q=\left|\dot{M}_{\mathrm{out}}/\dot{M}_{\mathrm{in}}\right|$) are
$0.8$, $0.58$, $0.8$ and $0.53$ for Models~I, II, III and IV. These
values indicate that relatively high efficiency of the outflow
production by the radiation for $\Gamma=0.6$. Interestingly, this
conversion efficiency $q$ (from the outflow to inflow) becomes smaller
for the models with gas rotations (Models~II and IV).

Figure~\ref{fig:mass-energy-flux} also shows the \emph{outflow}
powers ($P_{k}$ and $P_{\mathrm{th}}$) of the models as a function
of radius, as defined in eqs.~(\ref{eq:power-kinetic}) and (\ref{eq:power-thermal}).
As in the mass flux curves in the same figure, the dependency of the
energy flux on radius for the non-rotating cases (Models~I
and III) is almost identical to each other. Also for the rotating
cases (Models~II and IV), $P_{k}$ and $P_{\mathrm{th}}$ curves
are very similar to each others except for the small bumps around
$r'\sim10^{5}$ seen in the 2-D model (Model~II), but not in the
3-D model (Model~IV). The figure shows that in all four models, the
outflow power is dominated by kinetic process although the difference
between the kinetic power and the thermal power is much smaller than
in the models with gas rotation. In other words, the kinetic power
or the radiation force is more significant than the pressure gradient
force in these models. We also find that the kinetic powers at the
outer boundary dramatically decreases (more than an order of magnitude)
when the gas is rotating (Models~II and IV), but the thermal power
at the outer boundary dramatically increases when the gas is rotating
(cf., Tab.~\ref{tab:Model-Summary}). No significant difference
in the amount of $P_{k}$ and $P_{\mathrm{th}}$ between 2-D and 3-D
models is found. 

In summary, we find that the rotation reduces the outflow collimation,
and the outflow fluxes of mass and kinetic energy. Rotation also leads
to fragmentation and time variability of the outflow, but this effect
is reduced in the 3-D model (Model~IV) as the ring-like structure
seen in the 2-D model (Model~II) becomes distorted and the flow becomes
less organized. Rotation increases the outward flux of the thermal
energy also. Finally, the rotation does not change the mass inflow
rate through the outer boundary.


%


\subsection{Non-axisymmetric Nature of the Flows in 3-D }

\label{sub:Non-axisymetric-nature}

\begin{figure*}
\begin{center}

\includegraphics[clip,width=0.95\textwidth]{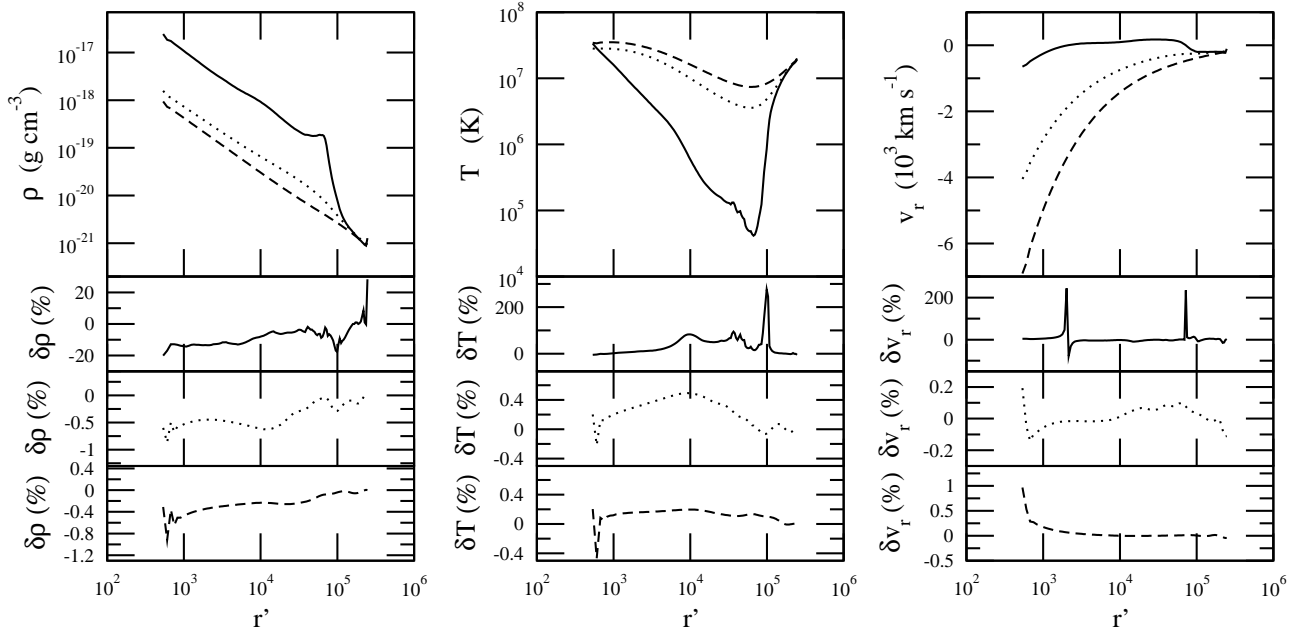}

\end{center}

\caption{Comparison of the density ($\rho$),
temperature ($T$) and the radial component of the velocity ($v_{r}$)
from the non-rotating gas models in 2-D (Model~I) and 3-D (Model~III).
The top panel shows the azimuthal angle averaged values of $\rho$,
$T$ and $v_{r}$ of Model~III along three different polar angles,
$\theta=5^{\circ}$~(\emph{solid line}), $45^{\circ}$~(\emph{dotted
line}) and $85^{\circ}$~(\emph{dashed line}), as a function of radius
($r'=r/r_{*}$). The lower three panels in each column show the percentage differences
between the azimuthal angle averaged values of the 3-D and the 2-D
models for each polar angle: $\theta=5^{\circ}$ (\emph{the second row}),
$\theta=45^{\circ}$ (\emph{the third row}), and $\theta=85^{\circ}$
(\emph{the fourth row}). The percentage difference values used here are
defined as $\delta x=(x_{\mathrm{3D}}-x_{\mathrm{2D}})\, x_{\mathrm{2D}}^{-1}\times100$\%
where $x$ is $\rho$, $T$ or $v_{r}$, and $x_{\mathrm{3D}}$ and
$x_{\mathrm{2D}}$ indicate the values for the 3-D and 2-D models
respectively. Along the relatively larger polar angles (i.e.,
$\theta=45^{\circ}$ and
$85^{\circ}$), little difference ($<1$~\%) is seen between the
models. The difference becomes much larger along $\theta=5^{\circ}$
line as it is very close to the outflow region in which the effect
of the radiative force due to line process is strongest. }

\label{fig:rho-temp-vr-Model-I-III}

\end{figure*}

\begin{figure*}
\begin{center}

\includegraphics[clip,width=0.95\textwidth]{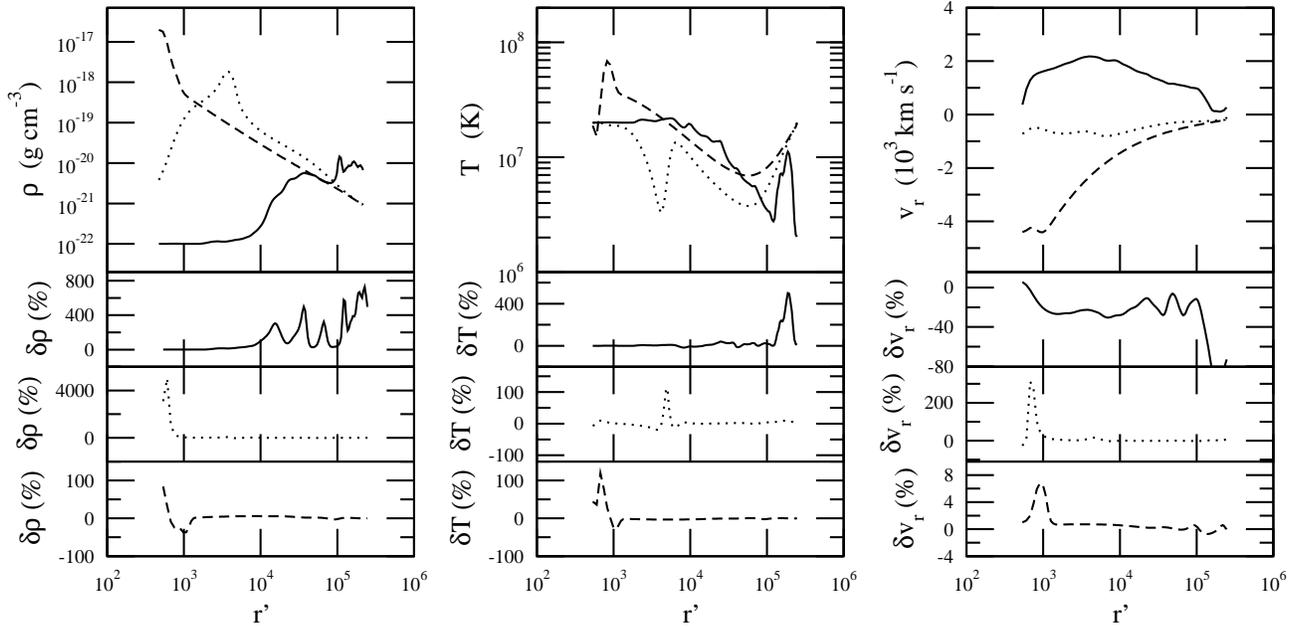}

\end{center}

\caption{As in Fig.~\ref{fig:rho-temp-vr-Model-I-III},
but for the rotating gas cases: Models~II and IV. Compared to the
non-rotating gas cases (Fig.~\ref{fig:rho-temp-vr-Model-I-III}),
the difference between the 2-D and 3-D models are larger since the
non-axisymmetric nature of flows in the 3-D model is more evident
in the rotating gas models (cf.~Figs.~\ref{fig:Density-3d} and \ref{fig:Density-Temp-Map-II-IV}). }

\label{fig:rho-temp-vr-Model-II-IV} 

\end{figure*}

Next, we compare the difference between the 2-D and 3-D models more
quantitatively. Figures~\ref{fig:rho-temp-vr-Model-I-III} and \ref{fig:rho-temp-vr-Model-II-IV}
show the gas density ($\rho$), temperature ($T$) and the radial
velocity ($v_{r}$) of the 3-D models with no gas rotation (Model~III)
and with gas rotation (Model~IV), respectively. The figures show that
values of $\rho$, $T$ and $v_{r}$ along three different polar angles
($\theta=5^{\circ}$, $45^{\circ}$, and $85^{\circ}$), but averaged
over azimuthal angle $\phi$, in order to compare the lines with those
of the 2-D models (Models~I and II, respectively). The figures also
show the percentage differences between the 2-D and 3-D models. 

For the non-rotating cases (Fig.~\ref{fig:rho-temp-vr-Model-I-III}),
the percentage differences of $\rho$, $T$ and $v_{r}$ between the
2-D and 3-D models are quite small ($<1$~\%) along relatively larger
polar angles, i.e., $\theta=45^{\circ}$ and $85^{\circ}$, indicating
the flow in along these lines are almost axi-symmetric. The difference
becomes much larger along $\theta=5^{\circ}$ line as it is very close
to the the region influenced by the outflow in which the effect of
the radiative force is strongest. 

As one can clearly see from the 3-D representation of the density
distribution (Fig.~\ref{fig:Density-3d}), the deviation from the
axisymmetry is much larger in the rotating cases. Figure~\ref{fig:rho-temp-vr-Model-II-IV}
shows that the percentage differences of $\rho$, $T$ and $v_{r}$
values between the 2-D and 3-D models (Models~II and IV) along the
three polar angles become very large ($>100$~\%) at some radii,
and they appear as sharp peaks or dips. These peaks and dips in the
percentage difference plots are caused by the presence of the cold cloud-like
structures which are stretched and drifted from the original ring-like
structures (as seen in the 2-D model, cf.~Fig.~\ref{fig:Density-3d}). 


%

\begin{figure*}
\begin{center}

\includegraphics[clip,width=0.95\textwidth]{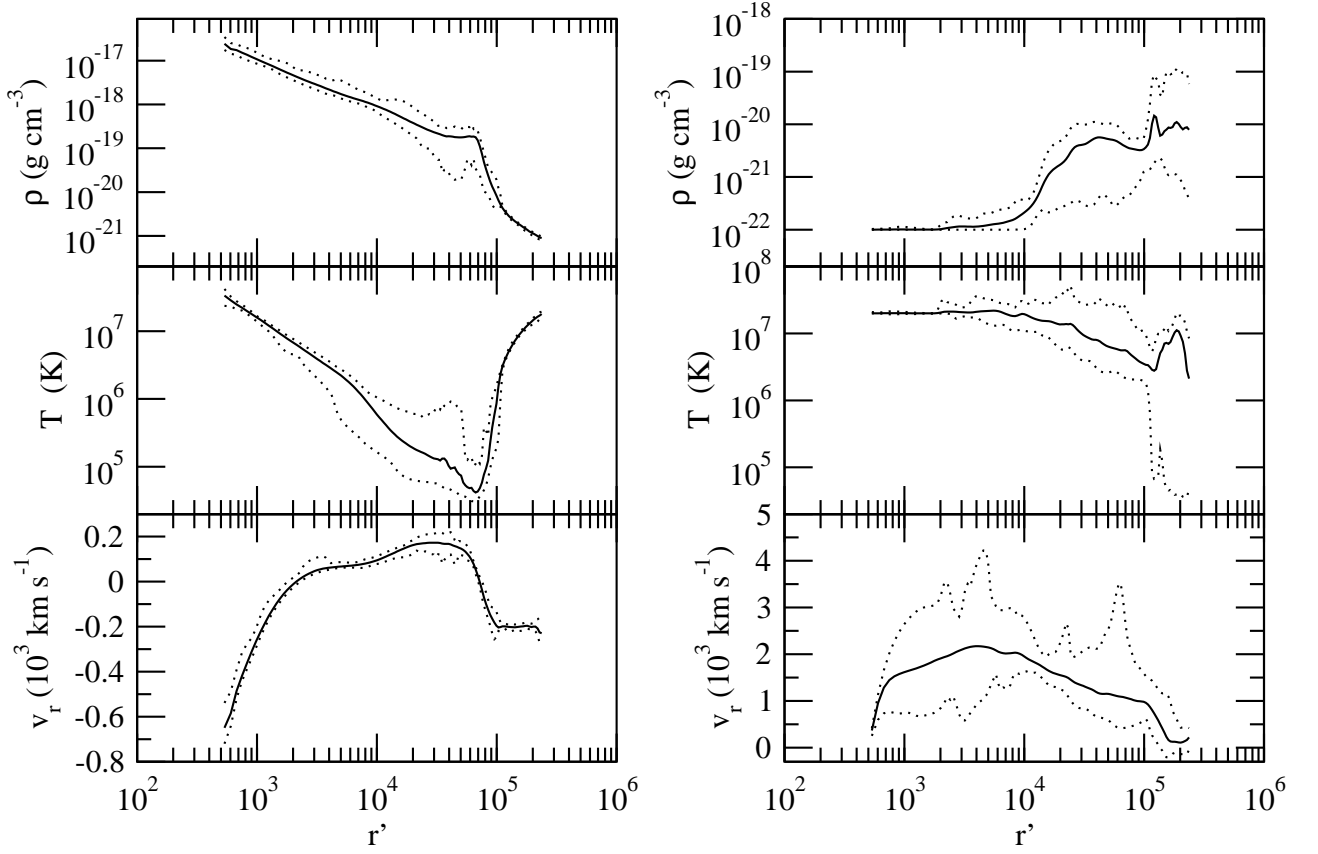}

\end{center}

\caption{The azimuthal angle ($\phi$) variations
of the density (\emph{top}), temperature (\emph{middle}) and radial
velocity (\emph{bottom}) along the polar angle $\theta=5^{\circ}$,
as a function of radius for the 3-D models: Models~III (\emph{left})
and IV (\emph{right}). Each panel shows the $\phi$ angle averaged
values (\emph{solid line}), the maximum values (\emph{upper dotted
line}) and the minimum values (\emph{lower dotted line}) around the
rotation axis. Both models
clearly show non-axisymmetric nature of the flows. The length scale
is in units of the inner disk radius ($r'=r/r_{*}$). }

\label{fig:phi-variation} 

\end{figure*}

To demonstrate the amount of azimuthal variations in density, temperature
and radial velocity in the 3-D models, we simply find their minimum
and maximum values around the symmetry axis ($z$-axis) for a fixed
polar angle $\theta$ as a function of radius, and compared them with
the azimuth angle averaged values. The results are shown in Figure~\ref{fig:phi-variation}
for the lines along the fixed polar angle of $\theta=5^{o}$. Both models
(Models~III and IV) show clear signs of azimuthal variation hence
the sings of non-axisymmetry at all radii. For the non-rotating case
(Model~III), the azimuthal variations of $\rho$, $T$ and $v_{r}$
are largest in a mid section ($r'=10^{4}$--$10^{5}$) while they
tend to increase as $r'$ increases for the rotating case (Model~IV),
except for that of $v_{r}$ which shows rather large variation at
all radii. The overall azimuthal variations of $\rho$, $T$ and $v_{r}$
in the rotating model are larger than those of the non-rotating model,
indicating that the degree of non-axisymmetry is larger for the
rotating case (Model~IV).  This is caused by the increase in the
amount of shear and thermal instabilities in the models with gas
rotation. 



\subsection{Properties of Gas --- Photoionization Parameter, Temperature, and
Radial Velocity}

\label{sub:Photoionization-paramters}

The volume averaged density ($\rho$) and temperature ($T$) of the
gas in all four models are about $2.2\times10^{-21}\,\mathrm{g\, cm^{-3}}$
and $1.4\times10^{7}\,\Kelvin$, and there is no significant difference
between the models. 
The volume averaged values of photoionization
parameters ($\xi$) are $1600$,
$1600$, $1600$, and $1500$ for Models~I, II, III, 
and IV respectively. Again, no significant difference between the
models is seen. As expected, the global properties of $\rho$, $T$ and $\xi$
seem to be mainly controlled by the outer boundary conditions ($T_{\mathrm{o}}=2\times10^{7}$~K
and $\rho_{\mathrm{o}}=1\times10^{-21}\,\mathrm{g\, cm^{-3}}$) and
the accretion luminosity, which are common to all the models presented
here. In the following, we examine the property of the gas in each
model more closely.

\begin{figure}

\begin{center}

\includegraphics[clip,width=0.48\textwidth]{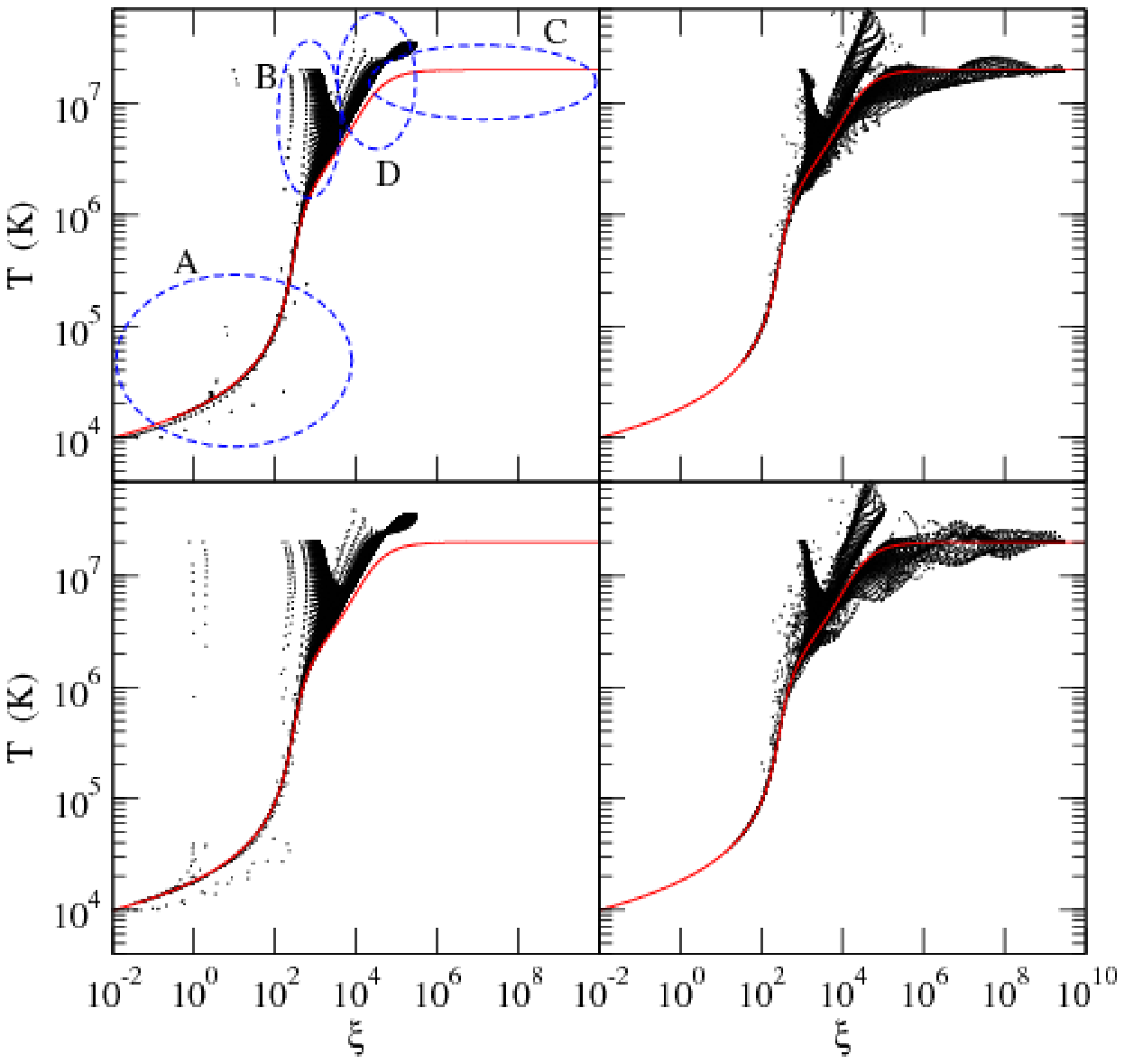}

\end{center}

\caption{Scatter plots of temperature ($T$) verses
photoionization parameter ($\xi$) from Models~I (\emph{upper left}),
II (\emph{upper right}), III (\emph{lower left}) and IV (\emph{lower
right}), overplotted with the cooling curve of the gas used in the
models (\emph{solid line}). To avoid overcrowding, only the points
on the $\phi=0$ plane are plotted for the 3-D models (\emph{lower panels}).
The gases from the 2-D and 3-D models occupy very similar phase spaces
for both non-rotating (\emph{left panels}) and rotating
(\emph{right panels}) cases. The $\xi$--$T$ planes are divided into four
distinctive regions (\emph{Regions~A, B, C} and \emph{D}), indicated
by the ellipses in the panel for Model~I. These regions apply to
all the models, but are not shown for clarity. }

\label{fig:phase-T-xi}

\end{figure}

The scatter plots of the temperature of the gas as a function of the
photoionization parameter $\xi$ for the models are shown in Figure~\ref{fig:phase-T-xi}
along with the cooling curve (assuming the radiative equilibrium)
used in our model [see eq.~(18) in \citealt{Proga:2000} or
  Paper~I]. For the 3-D models, only the points from the $\phi=0$ 
plane (cf., Figs.~\ref{fig:Density-Temp-Map-I-III} and \ref{fig:Density-Temp-Map-II-IV})
are shown in the figure to avoid over-crowding of the points. Although
the points from other $\phi$ planes are not shown here, by visual
inspections we find that the points shown here represent the distributions
of the whole samples. 

The figure shows that the overall distributions of the points on the
$\xi$--$T$ planes from the 2-D models are very similar to those
of the 3-D models. No significant difference between Models~I and
III is found, and neither between Models~II and IV. On the other
hand, the difference between the non-rotating cases (Models~I and
\textcolor{black}{III}) and the rotating cases (Models~II and IV) are clearly seen. The
 $\xi$--$T$ planes in the figures are divided into four main Regions
(A, B, C and D).  Although not shown here individually,
close inspections of the points, by separating them with different
ranges of $v_{r}$, $\rho$ and the distance from the central source
($r$), we found the following. 

\emph{Region~A}. The points in this region are mainly found in the
models without gas rotation (Models~I and III). The gas in this region
has relatively low temperatures ($T<10^{6}$~K), and has relatively low
values of photoionization parameter ($\xi<10^{2}$). They are found
at relatively small radii $r<0.5$~pc or equivalently $r'<1.8\times10^{4}$,
and have relatively large density ($\rho>10^{-20}\,\mathrm{g\, cm^{-3}}$).
They are \emph{outflowing} gas with relatively large radial velocities
($v_{r}>500\,\mathrm{km\, s^{-1}}$). 

\emph{Region~B}. The points in this region are found in both models
with (Models~I and III) and without (Models II and IV) gas rotation.
The temperature of the gas is relatively high ($T>10^{6}$~K), and
have median values of photoionization parameter ($\xi\sim10^{3}$).
They are found at relatively large distance from the center ($r>0.5$~pc),
and have relatively small density ($\rho<10^{-20}\,\mathrm{g\, cm^{-3}}$).
The gas in this region is \emph{mainly inflowing} with relatively
small radial velocities ($-500\,\mathrm{km\, s^{-1}}<v_{r}<0\,\mathrm{km\, s^{-1}}$). 

\emph{Region~C}. The points in this region are mainly found in the
models with rotations. The temperature of the gas is relatively high
($T \gtrsim 10^{7}$~K), and have relatively high values of photoionization
parameter ($\xi>10^{5}$). The points in this region are found at
relatively small radius ($r<0.5$~pc), and have relatively low density
($\rho<10^{-20}\,\mathrm{g\, cm^{-3}}$). The gas in this region is
\emph{outflowing} with relatively \textcolor{black}{large} radial velocity
($v_{r}>500\,\mathrm{km\, s^{-1}}$), and is found mainly near the rotation axis.
The property of the outflowing gas found here (in rotation cases)
is very different from that of the outflowing gas in the non-rotating
cases (\emph{Region~A}). 

\emph{Region~D}. The points in this regions are found in both non-rotating
and rotating cases, but a larger fraction of points are found in the
rotating cases. The temperature of the gas is relatively high ($T>10^{6}$~K),
and have median values of photoionization parameter ($\xi\sim10^{4}$).
The points in this region are found at relatively small radius ($r<0.5$~pc),
and have relatively high density ($\rho>10^{-20}\,\mathrm{g\, cm^{-3}}$).
The gas in this region is \emph{inflowing} with relatively \textcolor{black}{large}
radial velocity ($v_{r}<-500\,\mathrm{km\, s^{-1}}$).

From the close inspection of the different regions mentioned above,
we find that the deviations of the points on the $\xi$--$T$ plane
from the cooling curve are caused either by the compression/expansion
or by the outer boundary conditions. 
The points in Region~D, which are
found above the cooling curve, are over-heated by the compression
of the gas, as we found that the gas in this region is inflowing.
In Region~B, the gas is not in the radiative equilibrium because the
gas is located at large radii and its
thermal properties are influenced by the outer boundary condition, 
i.e.,  $T=2 \times 10^{7}$~K  regardless of $\xi$. 
Further, the points in Region~C, which are found in the outflow of
the rotating models and located mostly just below the cooling
curve, are slightly under-heated due to the influence of thermal expansion
of the gas. Lastly, we find that the points in Region A, which are
mainly in the non-rotating cases (Models~I and III), mostly follow
the cooling curve even though the points in regions are found to the
relatively high speed outflow. This is because the outflow
in the non-rotating models are mainly caused by the radiative pressure,
but not due to thermal expansion, as we found in the energy power
flux plot earlier in \S~\ref{sub:Mass-and-Energy} (Fig.~\ref{fig:mass-energy-flux})
whereas the thermal power is comparable to the kinetic power for the
rotating cases.



\begin{figure}

\begin{center}

\includegraphics[clip,width=0.48\textwidth]{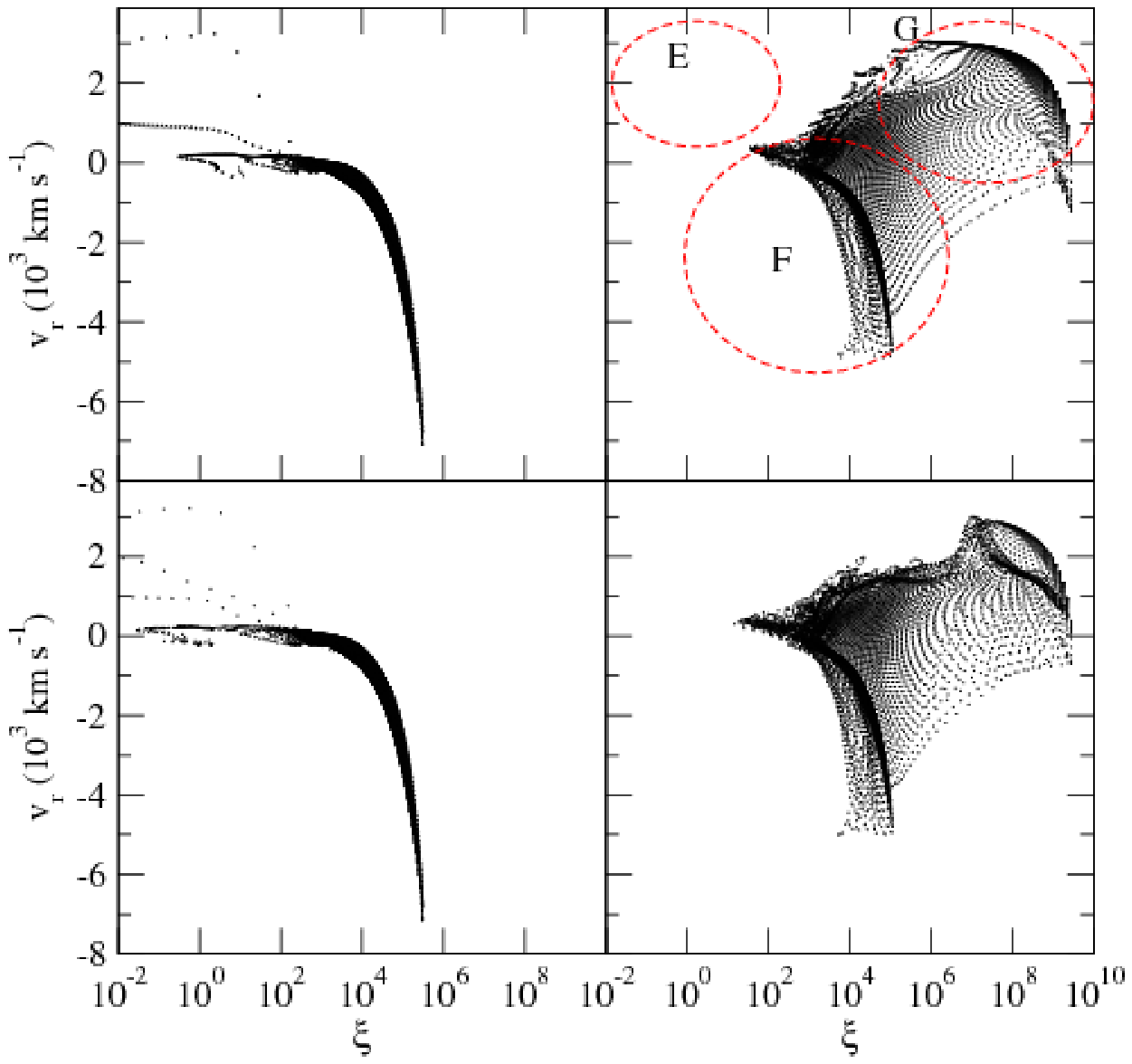}

\end{center}

\caption{Scatter plots of the radial velocity ($v_{r}$)
verses photoionization parameter ($\xi$) from Models~I (\emph{upper
left}), II (\emph{upper right}), III (\emph{lower left}) and IV (\emph{lower
right}). To avoid overcrowding, only the points on $\phi=0$ plane
are plotted for the 3-D models (\emph{lower panels}). The gases from
the 2-D and 3-D models occupy very similar phase spaces for both
non-rotating (\emph{left panels}) and rotating (\emph{right
panels}) cases. For the non-rotating cases, the majority of the outflowing
gas ($v_{r}>0$) has relatively 
low ionization parameter values ($\xi<10^{2}$), and no gas has
$\xi>10^{6}$.  A large fraction
of outflowing gas in the rotating cases has relatively high ionization
parameter values ($\xi>10^{6}$). The $v_{r}$--$\xi$ planes are
divided into three distinctive regions (\emph{Regions~E, F} and \emph{G}),
indicated by the ellipses in the panel for Model~II. These regions
apply to all the models, but are not shown for clarity. }

\label{fig:phase-vr-xi} 

\end{figure}

\begin{figure}

\begin{center}
\includegraphics[clip,width=0.48\textwidth]{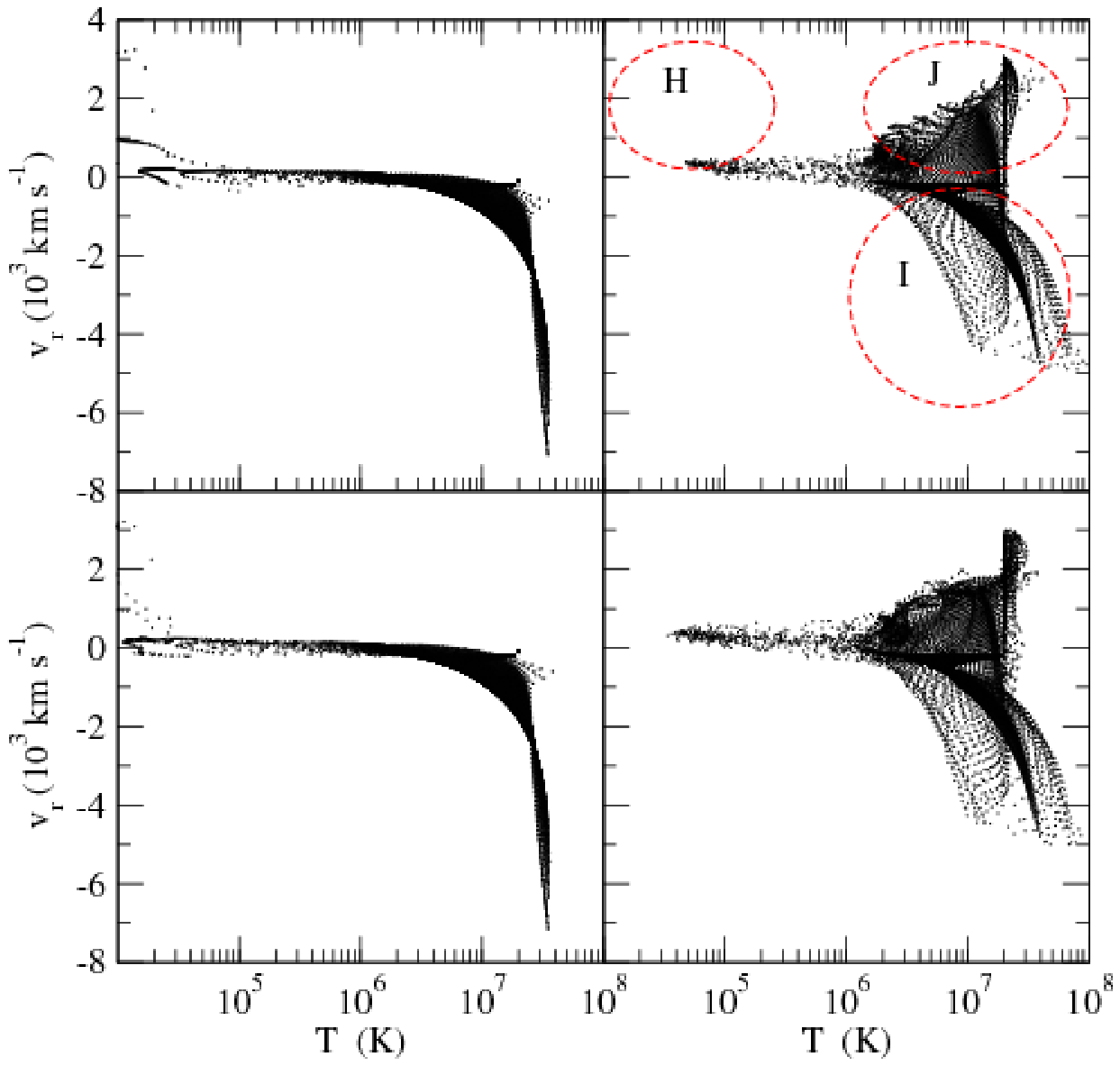}
\end{center}

\caption{Scatter plots of the radial velocity ($v_{r}$)
verses temperature ($T$) from Models~I (\emph{upper left}), II (\emph{upper
right}), III (\emph{lower left}) and IV (\emph{lower right}). To avoid
overcrowding, only the points on $\phi=0$ plane are plotted for the
3-D models (\emph{lower panels}). The gases from the 2-D and 3-D models
occupy very similar phase spaces for both non-rotating (\emph{left
panels}) and rotating (\emph{right panels}) cases. A large fraction
of gas is in outflow motion ($v_{r}>0$) for the models with rotation.
For the non-rotating cases (Models~I and III), the majority of the outflowing
gas has relatively low temperatures $T<10^{5}\,\Kelvin$ whilst a larger
range of the temperature is associated with the outflowing gas for
the rotating gas cases (Models~II and IV). See also the temperature
maps in Figs.~\ref{fig:Density-Temp-Map-I-III} and \ref{fig:Density-Temp-Map-II-IV}.
The $v_{r}$--$T$ planes are divided into three distinctive regions
(\emph{Regions~H, I} and \emph{J}), indicated by the ellipses in
the panel for Model~II. These regions apply to all the models, but
are not shown for clarity. }

\label{fig:phase-vr-temp}

\end{figure}

To see the difference in the properties of \textcolor{black}{the} outflowing gas between
the non-rotating and rotating cases, the scatter plots of $v_{r}$
vs $\xi$ and $v_{r}$ vs $T$ of the four models are shown in Figures~\ref{fig:phase-vr-xi}
and \ref{fig:phase-vr-temp}, respectively. \textcolor{black}{Both} $v_{r}$--$\xi$
and $v_{r}$--$T$ planes are divided into three distinctive regions
(Regions~E, F and G in Fig.~\ref{fig:phase-vr-xi}; Regions~H,
I and J in Fig.~\ref{fig:phase-vr-temp}). 

As in the previous $T$ vs $\xi$ scatter plots, the distribution
of the points are very similar between the 2-D and 3-D models. A small
difference between the 2-D and 3-D models is seen in Region~G
\textcolor{black}{(Fig.~\ref{fig:phase-vr-xi})} of 
the rotating cases. The points for the inflowing gas ($v_{r}<0$)
form a very similar pattern on the $v_{r}$--$\xi$ plane (Region~F
in Fig.~\ref{fig:phase-vr-xi}) for both rotating and non-rotating
cases. The largest inflow speed of the gas is slightly higher in the
non-rotating models, i.e., $v_{r}\sim-7000\,\mathrm{km\, s^{-1}}$
for the non-rotating models, and $v_{r}\sim-5000\,\mathrm{km\, s^{-1}}$
for the rotating models. A very noticeable difference between the
rotating and the non-rotating cases is seen in the outflowing gas
($v_{r}>0$). For the rotating models, the outflowing gas mainly appears
in Region~G where the photoionization parameter values are relatively
high ($\xi>10^{6}$) while for the non-rotating cases, it mainly
appears in Region~E where the photoionization parameter values are
relatively small ($\xi<10^{2}$). Again, this is due to the difference
in the dominating outflow mechanisms between the non-rotating and
the rotating cases, i.e., the outflow is mainly radiatively driven
for the non-rotating cases while the thermal pressure significantly
contributes to the outflows of the rotating cases (cf.~Fig.~\ref{fig:mass-energy-flux}). 



Rather similar patterns of the scattered points (to those in the $v_{r}-\xi$)
are seen in the $v_{r}$-- $T$ plane (Fig.~\ref{fig:phase-vr-temp}).
Again, the planes are divided into three regions (Regions~H, I and
J), and no significant difference between the distributions of the
points in the 2-D and the 3-D models is seen. The points for the
inflowing gas appear in Region~I in the rotating and the non-rotating
cases, and their distributions are somewhat similar to each other.
For the rotating models, the outflowing gas mainly appear in Region~J
where the gas temperatures are relatively high ($T>10^{6}$~K) while
for the non-rotating cases, they mainly appear in Region~H where
the temperatures are relatively small ($T<10^{5}$~K).

By comparing the physical properties of different regions in
Figures~\ref{fig:phase-T-xi}, \ref{fig:phase-vr-xi} and
\ref{fig:phase-vr-temp},  we found the following connections among 
them.  Regions~A, E and H are likely to belong to same grid
points (same spatial locations).  Region B corresponds to the upper 
section of Region~F.  The points in Regions~C, G and J are also likely
to belong to same grid points, so do the points in
\textcolor{black}{Regions~F and I}, respectively.



\section{Discussions}

\label{sec:Discussions}

\subsection{Virial Mass and Cold Clouds}

\label{sub:Virial-Mass}

To understand the evolution of galaxies which is greatly influenced
by the existence and the growth rate of the central SMBH, accurate
measurements of fundamental physical quantities such as mass of a SMBH
are important. While it is possible to estimate the masses directly
from the kinematics of the gas and stars for nearby systems, it is
difficult/impossible to apply this method for more distant objects
and for a very large number of objects (cf.~a review by \citealt{Ferrarese:2005}).
For the distant objects, the masses are estimated by the reverberation
mapping technique (cf.~a recent review by \citealt{Peterson:2006})
in conjunction with the virial theorem, i.e.,
\begin{equation}
  M_{\mathrm{BH}}=\frac{V^{2}R}{G}
  \label{eq:virial-theorem-original}
\end{equation}
where $V$ and $R$ are the average speed of an ensemble of the line
emitting clouds and the average distance of the ensemble of line emitting
clouds from the center. 

The mass estimate via the virial theorem uses the assumption that
the line emitting regions are gravitationally bounded and the outflows
are negligible. This assumption is not quite valid for the system
with relatively high Eddington number ($\Gamma=L/L_{\mathrm{Edd}}$),
as this is the case for our models ($\Gamma=0.6$). The outflow motions
of gas are clearly observed in our simulations too. In case of a point-source
approximation (for radiation source), the radiation force scales as
$r^{-2}$ (so does the gravitational force). Hence, the effective
gravity (including the radiation force term) will be reduced. Consequently,
the masses computed from the virial theorem will underestimate actual
masses, for the system with relatively large $\Gamma$. This effect
may be especially important for the Seyfert galaxies with high {[}\ion{O}{3}]~$\lambda$5007
blueshifts (``blue outliers'') which deviates from the $M_{\mathrm{BH}}$--$\sigma_{e}$
relation of normal, narrow-line Seyfert 1 (NLS1) and broad-line Seyfert~1
(BLS1) galaxies (\citealt{Komossa:2007}; \citealt{Komossa:2008}).
A recent work by \citet{Marconi:2008} explicitly demonstrates that
the correction for the virial mass estimate is significant when one
include the effect of radiation force (see also \citealt{Peterson:2000};
\citealt{Krolik:2001}; \citealt{Onken:2002}; \citealt{Collin:2006};
\citealt{Vestergaard:2006}). 

We apply the virial theorem to our simulation result to estimate the
BH mass in spite of the obvious outflows seen in our simulations,
and compare the value with the actual mass used in the simulation.
We restrict our discussion to the results of the 3-D model with gas
rotation (Model~IV). We assume the lines are formed in the
\textcolor{black}{dense} cold-cloud
like structures, which might resemble the narrow-line regions (NLR)
of AGN (found in \S~\ref{sub:density-temp-maps}). 
\textcolor{black}{The velocities and
positions of the cloud elements (the model grid points which belong
to the clouds) will be used in the virial theorem.} 
We define the gas to
be in \textcolor{black}{dense} cold-cloud state when its density is higher than
$\rho_{\mathrm{min}}=1.6\times10^{-20}\,\mathrm{g\, cm^{-3}}$ 
and its temperature is less than $T_{\mathrm{max}}=1.6\times10^{5}\,\mathrm{K}$. 

\begin{figure}
\begin{center}

\includegraphics[clip,width=0.48\textwidth]{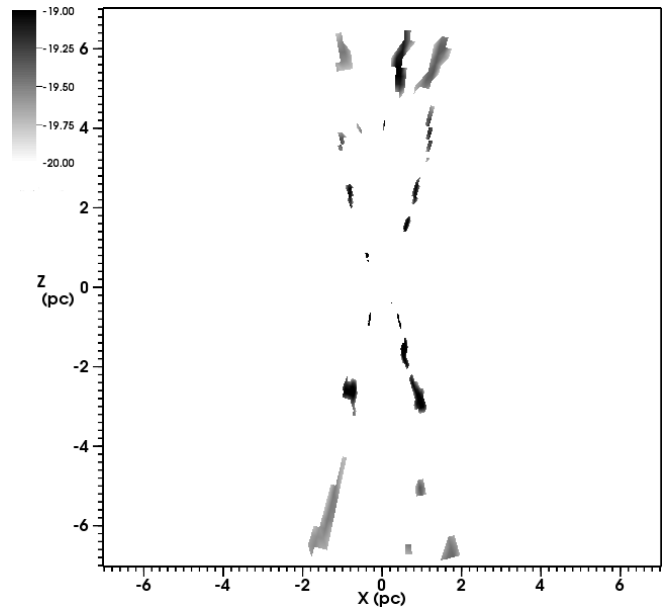} 

\end{center}

\caption{Spatial distributions of the {}``cold clouds''
in the 3-D model with gas rotation (Model~IV). The grayscale image
shows the density map of the cold clouds in logarithmic scale \textcolor{black}{(in cgs
unit)} on the
$z$--$x$ plane. The cold clouds here are defined as the gas with
its density higher than
$\rho_{\mathrm{min}}=1.6\times10^{-20}\,\mathrm{g\, cm^{-3}}$ and
its temperature less than $T_{\mathrm{max}}=1.6\times10^{5}\,\mathrm{K}$.
The clouds are not spherically distributed, but located near the bi-conic
surface (which appears as an X-shaped pattern here) defined by the outflowing
gas. Note that the length scale are in units of pc. }

\label{fig:cold-cloud}

\end{figure}

\begin{figure}
\begin{center}

\includegraphics[clip,width=0.48\textwidth]{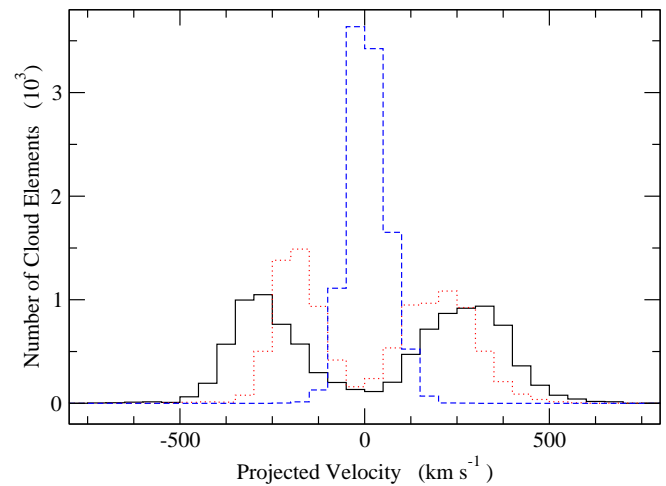} 

\end{center}

\caption{Histograms of the projected velocities ($v_{\mathrm{proj}}$)
of the cold cloud elements to an observer located at the inclination
angles ($i$) of $5^{\circ}$ (\emph{solid line}) , $45^{\circ}$(\emph{dotted
line}) and $85^{\circ}$(\emph{dashed line}). Note that an observer has
a pole-on view when $i=0^{\circ}$. While the distributions of $v_{\mathrm{proj}}$
for the lower inclination angles ($i=5^{\circ}$ and $45^{\circ}$)
show double peaks, that for the high inclination ($i=85^{\circ}$)
shows a single peak. This is caused by the bi-conic outflow morphology
of the cold clouds as seen in Figs.~\ref{fig:Density-Temp-Map-II-IV}
and \ref{fig:cold-cloud}. The separation between the double peaks
decreases as the inclination angle increases, as expected from the
bi-conic outflow morphology. }

\label{fig:v-projected} 

\end{figure}

Figure~\ref{fig:cold-cloud} shows the morphology of the cloud distribution
on the $z$--$x$ plane. The projected velocities ($v_{\mathrm{proj}}$)
of the cold cloud elements to an observer, located at the inclination
angles $i=5^{\circ}$, $45^{\circ}$ and $85^{\circ}$, are shown in Figure~\ref{fig:v-projected}.
The figure shows that the distributions of $v_{\mathrm{proj}}$ for
the lower inclination angles ($i=5^{\circ}$ and $45^{\circ}$) display
double peaks, and their separation decreases as the inclination angle
increases. These are expected features from the bi-conic outflow geometry
(as in Figs.~\ref{fig:Density-Temp-Map-II-IV} and \ref{fig:cold-cloud}). 

To compute the virial mass, we compute the average speed of the cold
cloud directly from our simulation result, i.e., $V=\left(\Sigma_{i=1}^{n}v_{i}\right)/n$
where $v_{i}$ and $n$ are the speed of an individual cold cloud
element and the total number of the clouds, respectively. Similarly,
the average radial distance is computed as $R=\left(\Sigma_{i=1}^{n}r_{i}\right)/n$
where $r_{i}$ is the radial distance of an individual cloud element.
For Model~IV, we find $V=285\,\mathrm{km\, s^{-1}}$
and $R=1.00 \times10^{19}\,\mathrm{cm}$. Note
that the escape velocity of the cold clouds, from the cloud forming
radius ($\sim1.5\times10^{4}\, r_{*}$) in Model~IV, is about $1.4\times10^{4}\,\mathrm{km\, s^{-1}}$
which is much larger than the average speed of the clouds ($V$).
The corresponding viral mass, using equation~(\ref{eq:virial-theorem-original}),
is $M_{\mathrm{vir}}=1.22\times10^{41}\,\mathrm{g}$
which is about $40$~\% smaller than the actual mass of the BH used
in the simulation, i.e., $M_{\mathrm{BH}}=1.989\times10^{41}\,\mathrm{g}$.
This is in general agreement with the previous statement: the virial
mass determined using equation~(\ref{eq:virial-theorem-original})
would underestimate actual mass for systems with relatively high $\Gamma$
in which the radiation force is comparable to or greater than the gravitational
force. A systematic correction for the radiation force in the virial
mass estimate, in general, is very challenging since the radiation
force (line force) depends on the ionization state of the gas, and
its strength is not spherically symmetric. Further, the outflow geometry
is non-spherical, and it depends on the rotation rate of the gas (cf.~Models
III and IV in Fig.~\ref{fig:Density-3d}). 



%


\subsection{Comparisons with Observations of Seyfert Galaxies}

\label{sub:Comparisons-with-NLS1}

The \textcolor{black}{studies of kinematics in} the NLR of Seyfert galaxies will provide
us a hint for understanding the complicated dynamical processes and
the driving forces (radiation, magnetic or thermal) in their vicinity.
The NLR of nearby Seyfert galaxies are especially useful for testing
outflow models since they can be spatially resolved (e.g., \citealt{evans:1993};
\citealt{macchetto:1994}; \citealt{hutchings:1998}; \citealt{nelson_c:2000};
\citealt{crenshaw:2000}; \citealt{crenshaw:2000b}; \citealt{ruiz:2001};
\citealt{cecil:2002}; \citealt{Ruiz:2005}; \citealt{das:2005, das:2006};
\citealt{kraemer:2008}; \citealt{Walsh:2008}). In particular, the Faint Object Camera (FOC)
and the Space Telescope Imaging Spectrograph (STIS) on \emph{HST},
allow for detailed constraints on the kinematics of the NLR in Seyfert
galaxies. For example, using the STIS,  \citet{das:2005} obtained the
position dependent spectra of {[}\ion{O}{3}]~$\lambda5007$ for NGC~4151,
one of the closest Seyfert galaxies, with different long slit positions,
and studied the kinematics of the wind in the NLR by measuring its
projected velocity components from the position of multiple peaks
(up to three peaks) in the {[}\ion{O}{3}] profiles. Their results are
very intriguing. For scales from 10~pc to 100~pc, they found that the velocity increases
nearly linearly with radius whereas at larger scales, the velocity
decreases, again nearly linearly, with increasing radius. Spatially
resolved observations of the NLR in other AGN show similar flow patterns
(e.g., NGC~1068: \citealt{crenshaw:2000}; \citealt{kraemer:2000b}
and Mrk~3: \citealt{Ruiz:2005}).

To compare our model with the kinematics study of NGC~4151 \citet{das:2005},
we compute the velocity of the cold clouds (as defined in \S~\ref{sub:Virial-Mass})
in Model~IV (cf.~Fig.~\ref{fig:cold-cloud}) projected ($v_{\mathrm{proj}}$)
toward an observer at the inclination angle $i=45^{\circ}$, which
is also \textcolor{black}{the} inclination of NGC~4151 (\citealt{das:2005}). \citet{das:2005}
used that the kinematics model of the outflows with a bi-conic radial
velocity law, and found a good fit to their observations 
when the opening angle of the cone is $\sim 33^{\circ}$. Interestingly,
we find the opening angle of the outflows in Model~IV is also about
$30^{\circ}$ (cf.~Figs.~\ref{fig:Density-3d} and \ref{fig:Density-Temp-Map-II-IV}). 

\begin{figure}
\begin{center}

\includegraphics[clip,width=0.48\textwidth]{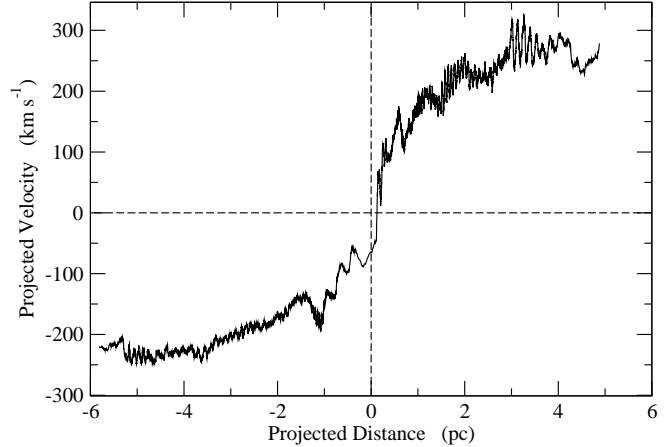} 

\end{center}

\caption{The velocities of the cold cloud elements
(as in Fig.~\ref{fig:cold-cloud}) projected toward an observer located
at the inclination angle $i=45^{\circ}$ are shown as a function of
the projected vertical distance (the distance along the $z$-axis
in Fig.~\ref{fig:cold-cloud}, but projected on to the plane of the
sky for the observer viewing the system with $i=45^{\circ}$). The
negative projected distance indicates the clouds are found in the
lower half of the projection plane. The clouds are accelerated up
to $\sim4$~pc, but the velocity curve flattens beyond this point.
Towards the outer edges (near the outer boundaries), the curve shows a sign
of deceleration. Although in different scales, the flow pattern resembles
the outflow kinematics of the NLR in Seyfert Galaxy NGC~4151 by \citet{das:2005}. }

\label{fig:v-projected-das} 

\end{figure}

Figure~\ref{fig:v-projected-das} shows $v_{\mathrm{proj}}$ of the
clouds plotted as a function of the projected vertical distance, which
is the distance along the $z$-axis in Fig.~\ref{fig:cold-cloud}
projected onto the plane of the sky for an observer viewing the
system with $i=45^{\circ}$. The figure shows that the clouds are
accelerated up to $250\,\mathrm{km\, s^{-1}}$ until the projected
distance reaches $\sim4$~pc, but the velocity curve starts to flatten
beyond this point. Towards the outer edges (near the outer boundaries),
the curve begins to show a sign of deceleration, but not so clearly.
We note that the hot outflowing
gas, on the other hand, does show deceleration at the larger radii
in our models (cf.~Fig.~\ref{fig:phi-variation}).
Although the physical size of the long slit observation of NGC~4151
by \citet{das:2005} is in much lager scale ($\sim50$ times larger)
than that of our model, their radial velocities as a function of the
position along the slit (see their Figs.~5 and 6) show a similar
pattern as in our model (Fig\@.~\ref{fig:v-projected-das}). The
range of $v_{\mathrm{proj}}$ in our model is about $-250$ to $300\,\mathrm{km\, s^{-1}}$
while the range of the observed radial velocities in \citet{das:2005}
is about $-800$ to $800\,\mathrm{km\, s^{-1}}$, which is comparable
to ours. To understand the large scale outflows seen in the observations
and to understand the kinematics of such outflows better, the size
of the simulation box must be increased at least by a factor of
100. In such larger scales, the temperature is expected to be much cooler,
and the dust would play an important role  in determining the thermal
and dynamical properties the outflows (e.g.~\citealt{Antonucci:1984};
\citealt{Miller:1990}; \citealt{Awaki:1991}; \citealt{Blanco:1990};
\citealt{Krolik:1999}).  
These are beyond the scope
of this paper, but shall be considered in a future paper.


\section{Conclusions}

\label{sec:Conclusions}

We have presented the dynamics of gas under the influences of
the gravity of a SMBH and the radiation force from the luminous accretion
disk around the SMBH. This is a direct extension of the previous axi-symmetric
models of Paper~I and Paper~II to a full 3-D
model, and is an extended version of the models presented in \citet{Kurosawa:2008}
to which we have added the radiation force due to line processes and
the radiative cooling and heating effect. We have considered two cases
from Paper~I and Paper~II:
(1)~the formation of outflow  from the accretion of the ambient
gas with no rotation and (2)~that with \textcolor{black}{weak} rotation. The models have
been considered in both 2-D and 3-D hence, in total, four models have
been presented. Our first main goal is to examine
if there is a significant difference between two models with
 identical initial and outer boundary conditions but in different dimensionality (2-D and 3-D). In
particular, we examine whether the radiation driven outflows that were found 
to be stable in the previous studies in 2-D (Paper~I; Paper~II)
still remain stable in 3-D. Our second main goal is to gain some
insights into the gas dynamics in AGNs and Seyfert galaxies by comparing
the simulation results with observations. In the following, we
summarize our main findings through this investigation.

1.~For non-rotating gas cases, the outflow occurs in very narrow
cones (with the opening angles $\sim5^{\circ}$) in polar directions.
Overall density and temperature of the both 2-D and 3-D models (Models~I
and III) are very similar to each other (Figs.~\ref{fig:Density-3d}
and \ref{fig:Density-Temp-Map-I-III}). Small but noticeable differences
are seen in the narrow outflow regions. 

2.~Rotation of gas significantly changes the morphology of the outflows
(Models~II and IV in Figs.~\ref{fig:Density-3d} and \ref{fig:Density-Temp-Map-II-IV}).
The centrifugal force pushes the outflow away from the polar axis
and forms much wider outflows (with the opening angles $\sim30^{\circ}$).
The outflow occurs mainly on and near bi-conic surfaces, and relatively
low values of density are found in the polar directions, unlike the
outflows in the non-rotating cases. The models with gas rotation
show cold clouds (clumps) in their outflows in their
2-D density and temperature maps (Fig.~\ref{fig:Density-Temp-Map-II-IV}).
Although the overall density and temperature structures of the flows
of the 2-D and 3-D models are similar to each other, the outflows
in 3-D occur in much less organized manner. We find that the cloud-like
structures seen in the 2-D model (Model~II), which are rings if the
density is expanded in 3-D using the axisymmetry 
(Fig.~\ref{fig:Density-3d}), are not stable in full 3-D simulations
due to the shear and thermal instabilities. 
The rings break up into smaller pieces, and fully 3-D clouds are
formed in Model~IV.

3.~The mass and energy fluxes plotted as a function of radius for
the 3-D non-rotating case are almost identical to those of the non-rotating
2-D case (Fig.~\ref{fig:mass-energy-flux}). For the rotating cases,
the bumps seen in the mass-inflow rate and the net mass flux at the
outer radii ($r' \gtrsim 10^{4}$) for the 2-D model (Model II) are smoothed
out in the 3-D model (Model~IV) due to the fragmentation of the ring
structures in the 3-D model. While the kinetic power dominates at
all radii for the non-rotating cases, the thermal power contributes
significantly to the outflow driving force for the rotating cases.
In spite of the differences in the flow geometries, the rotating models
in both 2-D and 3-D show very similar values of the mass accretion
and outflow rates at the outer and inner boundaries
(Table~\ref{tab:Model-Summary}). 
In other words, AGN feedback due to radiation is similar
in the 2-D and 3-D cases as far as the time-averaged mass and energy fluxes 
are concerned.

4.~For the non-rotating cases, the amount of variability in the mass flux
at the outer boundary is higher in the 2-D model than that in the
3-D model, but the opposite is true for the rotating cases (Fig.~\ref{fig:mdot-evolution}).

5.~\textcolor{black}{In the 3-D models}, the deviations from the axisymmetry are observed in
both rotating and non-rotating cases (Figs.~\ref{fig:rho-temp-vr-Model-I-III}
and \ref{fig:rho-temp-vr-Model-II-IV}). The amounts of the azimuthal
angle variations of the density, temperature, and radial velocity
(Fig.~\ref{fig:phi-variation}) are relatively small for the non-rotating
case (Model~III), but they are relatively large at all radii for
the rotating case (Model~IV). 

6.~The gas properties of the 2-D and 3-D models are very similar
to each other for both non-rotating and rotating cases (Figs.~\ref{fig:phase-T-xi},
\ref{fig:phase-vr-xi} and \ref{fig:phase-vr-temp}). The majority
of the outflowing gas for the rotating cases (Models~II and IV) has
relatively large values of the photoionization parameter ($\xi>10^{6}$)
while for the non-rotating cases, it has relatively small values of
the photoionization parameters ($\xi<10^{2}$)
(Fig.~\ref{fig:phase-vr-xi}). This is due to the difference in the dominant outflow mechanisms
between the non-rotating and the rotating cases, i.e., the outflow
is mainly radiatively driven for the non-rotating cases while the
thermal pressure significantly contributes to the outflows of the
rotating cases (cf., Fig.~\ref{fig:mass-energy-flux}). For the rotating
models, the majority of the outflowing gas has relatively high ($T>10^{6}$~K)
temperature while for the non-rotating cases, it has relatively low
($T<10^{5}$~K) temperature. The higher $\xi$ values seen in the
rotating cases are mainly from the \textcolor{black}{low-density} hot outflowing gas in between the
outflowing cold clouds. 

7.~For Model~IV, we find the average speed and the radial position
of the cold cloud elements (\S~\ref{sub:Virial-Mass}) as $V=285\,\mathrm{km\, s^{-1}}$
and $R=1.00\times10^{19}\,\mathrm{cm}$. The corresponding
viral mass is $M_{\mathrm{vir}}=1.22 \times10^{41}\,\mathrm{g}$
which is about 40~\% smaller than the actual mass of the BH used
in the simulation, i.e., $M_{\mathrm{BH}}=1.989\times10^{41}\,\mathrm{g}$.
This is in general agreement with the previous studies (e.g.~\citealt{Peterson:2000};
\citealt{Krolik:2001}; \citealt{Marconi:2008}) which predict that
the virial mass estimated without considering the effect of the radiation
force underestimates the actual mass of the SMBH. 

8.~The opening angles ($\sim30^{\circ}$) of the bi-conic outflows
found in the the rotating models (Models~II and IV) are very similar
to that of the nearby Seyfert galaxy NGC~4151 ($33^{\circ}$) determined
by \citet{das:2005}. Although the physical size of the long slit
observations of NGC~4151 by \citet{das:2005} is in much lager scale
($\sim50$ times larger) than that of our model, their radial velocities
as a function of the position along the slit (see their Figs.~5 and
6) show a similar pattern as in our model (Fig\@.~\ref{fig:v-projected-das}).
An important difference between the observation of \citet{das:2005}
and our models is the lack of clearly decelerating clouds at larger
radii in our models. However, we note that the clouds found in our
simulations reach a constant velocity near the outer boundary of our
simulations, and show a hint of deceleration. This puzzling outflow
deceleration seen in the observations might be due to the inflow that
interacts with the polar outflows. The reason for no clear
cloud deceleration seen in our model may be simply due to the
relatively small simulation box size we used, and the issue could be
resolved in a lager scale 
simulation. Spectroscopic studies of the NLR of Seyfert galaxies by
\citet{Komossa:2008} also favor a scenario in which the NLR clouds
are traveling in decelerating wind.
The hot outflowing gas, on the
other hand, does show deceleration at the larger radii 
in our models (cf.~Fig.~\ref{fig:phi-variation}).

To perform a better comparison of our models with observations hence to
constrain the model parameters, in future studies, we need to increase the size of the simulation
box to match the physical sizes of the NLR of Seyfert galaxies. It would
take the outer radius of the computational domain to be expanded by 
one or even two orders of magnitude compared to the one used here.
The dust is very likely important in the dynamics of the outflow in the larger
scale simulations since the temperature becomes low enough for the
dust survival and formation in the larger radius. 
We showed in Paper~II, relatively high density set at the outer boundary
promotes formation of cold clouds. Therefore, we plan to explore the effects
of dust and \textcolor{black}{outer boundary} density. 

To compare the model results
directly with observations, we would need to compute the radiative transfer
models of the important emission lines (e.g., {[}\ion{O}{3}]~$\lambda5007$,
H$\beta$ and \ion{C}{4}~$\lambda1549$), which will be the topic of our
future paper. Taking these steps will allow a quite
strict test of our results against observations of Seyfert galaxies and AGN.
It would also be interesting to check if our models could reproduce 
large-scale outflows in quasars, for example
the high-velocity outflow components seen in \ion{C}{4} and \ion{Mg}{2}
quasar absorption-line systems (e.g., see a recent work by
\citealt{Wild:2008}) which \textcolor{black}{would} provide an
additional \textcolor{black}{constraint} on our wind models.

\acknowledgements{Authors thank the anonymous referee for constructive
  comments and suggestions for improving the clarity of the
  manuscript.  This work was supported by NASA through grant
  HST-AR-11276 from the Space Telescope Science Institute, which is
  operated by the Association of Universities for Research in
  Astronomy, Inc., under NASA contract NAS5-26555. A significant
  fraction of our simulations was performed on a SUN computer system
  funded by President of UNLV, D. B. Ashley through an Infrastructure
  Award to the Astronomy Group at UNLV.  This work was also supported
  by the National Center for Supercomputing Applications under
  AST070036N which granted the accesses to the Xeon Linux Cluster
  Tungsten and Intel 64 Linux Cluster Abe.  Authors are grateful for
  the original developers of ZEUS-MP for making the code publicly
  available. }


\newpage





\newpage

\newpage

\newpage

\newpage

\newpage

\newpage

\newpage

\newpage

\newpage

\newpage

\newpage

\newpage

\newpage

\end{document}